\documentclass{article}

\usepackage{graphicx}  
\usepackage{amsmath}   
\usepackage{cite}
\usepackage{amssymb}   
\usepackage{bm} 
\usepackage{dcolumn}
\usepackage{color}
\usepackage{mathrsfs}
\usepackage{amsfonts}
\usepackage{varioref}
\usepackage{braket}
\usepackage{bm}
\usepackage[colorlinks,citecolor=blue,urlcolor=magenta,linkcolor=blue]{hyperref}

\labelformat{section}{Section #1} 
\labelformat{subsection}{Section #1} 
\labelformat{subsubsection}{Section #1}
\labelformat{subsubsubsection}{Section #1}
\labelformat{equation}{Eq.~(#1)} 
\labelformat{figure}{Fig.~#1} 
\labelformat{subfigure}{Fig.~\thefigure#1} 
\labelformat{table}{Tab.~#1} 
\labelformat{appendix}{Appendix #1}

\title{Inverting a normal harmonic oscillator: Physical interpretation and applications}
\author{Karthik Rajeev\footnote{karthik@iucaa.in}$~^{1}$,~ 
	Sumanta Chakraborty\footnote{sumantac.physics@gmail.com}$~^{2}$~and~
	T. Padmanabhan\footnote{paddy@iucaa.in}$~^{1}$\\
	{\small{$^{1}$IUCAA, Post Bag 4, Ganeshkhind, Pune University Campus, Pune 411007, India}}\\
	{\small{$^{2}$Department of Theoretical Physics, Indian Association for the Cultivation of Science, Kolkata-700032, India}}}

\begin{document}
	
	\maketitle
	
	\begin{abstract}		
		A harmonic oscillator with time-dependent mass $m(t)$ and a time-dependent (squared) frequency $\omega^2(t)$ occurs in the modelling of several physical systems. It is generally believed that  systems, with $m(t)>0$ and $\omega^2(t)>0$ (normal oscillator) are stable  while systems with $m(t)>0$ and $\omega^2(t)<0$ (inverted oscillator) are unstable. We show that it is possible to represent the \textit{same} physical system either as a normal oscillator or as an inverted oscillator by redefinition of dynamical variables. While we expect the physics to be invariant under such redefinitions, it is not obvious how this invariance actually comes about. We study the relation between these two, normal and inverted,  representations of an oscillator in detail both in Heisenberg and Schr\"{o}dinger pictures to clarify several conceptual and technical issues. The situation becomes more involved when the oscillator is coupled to another (semi)classical degree of freedom $C(t)$ and we want to study the back-reaction of the quantum system $q(t)$ on $C(t)$,  in the semi-classical approximation. We provide a simple prescription for the back-reaction based on energy conservation and study that the dynamics of the full system in both normal and the inverted oscillator representation. The physics again remains invariant but there are some extra subtleties which we clarify.  The implications of these results for quantum field theory in cosmological backgrounds are discussed briefly in an appendix.
	\end{abstract}

    \section{Introduction}
    
    Harmonic oscillators with time-dependent mass $m(t)$ and squared frequency $\omega^2(t)$ are ubiquitous in physics. They represent simplest non-trivial physical systems influenced by an external background which usually provides the time-dependence in $m(t)$ and $\omega(t)$. It is generally believed that such systems (with $m(t)>0$) are stable when $\omega^2(t) >0$. In this case, the Hamiltonian, though time dependent, is bounded from below. On the other hand, if $m(t)>0$ but $\omega^2(t)<0$ the Hamiltonian is \textit{not} bounded from below and one would suspect that such a system is inherently unstable. Based on this consideration, one would have also thought that the coupling of $q(t)$ to another degree of freedom, say $C(t)$, will induce a very different kind of behaviour in $C(t)$ depending on whether $q(t)$ is a normal oscillator or an inverted oscillator. This will be the case even when we treat $q(t)$ as a quantum mechanical system coupled to a semi-classical system $C(t)$ with the  physics described by some suitable semi-classical approximation. 
    
    A surprising fact is that one can  ``invert'' a normal oscillator by redefining the dynamical variable and, if necessary, the time coordinate. That is, the same physical system can be represented either as a normal oscillator or as an inverted oscillator depending on  the variables chosen to describe the system. Since such redefinition of dynamical variables and time coordinate should not affect the physics, we will expect both the normal and the inverted oscillator representations to have the same observable consequences. However, it is not a priori clear how this result actually comes about. One of the main goals of this paper is to study the transformation from the normal to inverted oscillator and describe the physical situation in both Schr\"{o}dinger and Heisenberg picture thereby clarifying several conceptual and technical issues. This is done in the first part of the paper.
    
    The situation becomes more interesting (and complicated) when we probe the system at a deeper level. In several physical contexts the time dependence of the mass $m(t)$ and frequency $\omega(t)$ of our system arises because it is coupled to some external agency. One simple example of such a situation is described by a Lagrangian of the following kind: 
    \begin{equation}
     L[q(t),C(t)] = \frac{1}{2} m(C)[ \dot q^2 - \omega^2(C) q^2 ] + \frac{1}{2} M [\dot C^2 - V(C)]
    \end{equation} 
    where $M$ is a constant.
    This Lagrangian describes a coupled system made of two degrees of freedom $q(t)$ and $C(t)$ and the coupling arises through the functions $m(C)$ and $\omega(C)$. If we think of $C$ as ``heavy'' degree of freedom and $q$ as a ``light'' degree of freedom (which is the usual parlance used, for e.g., in molecular dynamics) then, to the lowest order, we can ignore the influence of $q$ on $C$. In this, zeroth order approximation, the dynamics of $C$ can be taken to be classical and its evolution is described by a suitable solution to the equation of motion $\ddot C + V'(C)\approx0$. In this approximation we can think of $m[C(t)] \equiv m(t)$ and $\omega[C(t)] \equiv \omega(t)$ as specified functions of time determined by the evolution of $C(t)$. It is then possible to consider the quantum mechanics of the $q$ degree of freedom by treating as an oscillator with time dependent mass and frequency. 
    
    At the next order of approximation one would like to consider the back-reaction of the quantum degree of freedom $q(t)$ on $C(t)$ and obtain the lowest order corrections to the equation $\ddot C + V'(C)\approx0$ with some back-reaction force $B(C)$ appearing on the right hand side. This is the domain of semi-classical approximation in which one is considering a hybrid quantum-classical systems. The fact that the $q$-degree of freedom can be represented either as a normal oscillator or as an inverted oscillator introduces a new twist at this stage, viz., when we  quantize the oscillator degree of freedom $q(t)$ in the external background provided by $C(t)$ \textit{and also} take into account the back-reaction of the quantum oscillator on the semi-classical dynamics of $C(t)$. When we change the representation for the oscillator from the normal one to the inverted one, one might think that the back-reaction will get altered because of the ``instability'' of the inverted oscillator. But, again, since this inversion arises due to redefinition of the dynamical variable and the time coordinate, we do not expect the physical consequences to change under such a transformation. To demonstrate this, we need a consistent prescription for semi-classical back-reaction which can be obtained from demanding the conservation of total energy. Because of the time dependence experienced by the $q$-degree of freedom, due to the background evolution of $C(t)$, the energy of the $q$-mode will not be conserved. This energy has to be supplied by the $C$-degree of freedom. The correction due to back-reaction should ensure that the work done by the extra back-reaction force $B(C)$ correctly accounts for the energy transfer from $C$ to $q$. As we shall see, this criterion is powerful enough to provide a natural prescription for the back-reaction of $q$ on $C$. We use this prescription  to investigate what happens when we switch from the normal to the inverted oscillator. We  again expect that such a change of representation should leave the physical consequences unchanged.  In the second part of the paper we address this issue in detail and show that physics \textit{does} remain invariant, if we switch from normal to the inverted representation for $q$, even when we take into account the back-reaction of $q$ on $C$. This, however, comes about in a rather subtle manner and, for example, involves dealing with Lagrangians containing dependences on $\ddot C$. We explore and clarify these subtleties.
    
    One of the physical systems widely discussed in literature, which can be mapped to a bunch of harmonic oscillators with time dependent mass and frequency, is provided by the dynamical evolution of a scalar field in an expanding Friedman background. When we ignore back-reaction one can find a direct mathematical correspondence between such a system and the model discussed in the first part of the paper. The inclusion of back-reaction of a quantum field on classical gravity is a non-trivial issue because of well known divergences and ambiguities. In view of this we confine our discussion to systems with finite number of degrees of freedom in the main body of the paper; we have, however, discussed briefly the correspondence between our model and QFT in an expanding background along with some of the relevant caveats in \ref{Appendix_A}.

      \section{Turning an oscillator upside-down}\label{inverted_oscillator}
    
   Let us start with a harmonic oscillator Lagrangian with $m(t)>0$ and $\omega^2(t)>0$ and see how it can be transformed to an inverted oscillator with unit mass and imaginary frequency.\footnote{As a special case, one can also make the new frequency a constant, which we will comment about in the last section and discuss in a separate publication \cite{bkreactn}.} Since the implications of this feature --- viz., that the same physical system can be represented either as an oscillator with $\omega^2(t)>0$ or as an oscillator with $\omega^2(t)<0$ --- do not seem to have received adequate attention in the literature, let us examine it in some detail. 
    
    To do this, let us start with a `normal' oscillator with time-dependent mass $m(t)>0$ and (squared) frequency $\omega^2(t)>0$ described by the action:
    \begin{align}\label{actiontimedependent}
    \mathcal{A}=\frac{1}{2}\int dt~ m(t)\left[\dot{q}^{2}-\omega ^{2}(t)q^{2}\right]
    \end{align}
    This system has a Hamiltonian which is bounded from below for all $t$. Let us now change the dynamical variable from $q$ to $Q=\sqrt{m(t)}\, q$. This is a perfectly valid change of variables connecting a real $q$ to a real $Q$, since, by assumption, $m(t)$ is positive for all $t$. In terms of $Q(t)$, the action becomes: 
    \begin{align}
    \label{action_negative_freq}
    \mathcal{A}=\frac{1}{2}\int dt~ \left[\dot{Q}^{2}-\Omega ^{2}(t)q^{2}\right]-\frac{1}{4}\left(\frac{\dot{m}Q^2}{m}\right)\bigg\vert_{t_1}^{t_2} \rightarrow \frac{1}{2}\int dt~ \left[\dot{Q}^{2}-\Omega ^{2}(t)q^{2}\right]
    \end{align}
    where (with $g=\sqrt{m}$),
    \begin{align}
    \label{mass_sqrt_frequency}
    \Omega^2(t)=\omega^2(t)-\frac{\ddot{g}}{g}
    =\omega^2(t)+\frac{\dot{m}^2}{4m^2}-\frac{\ddot{m}}{2m}
    \end{align}     
    In the second part of \ref{action_negative_freq} we have dropped  the boundary term in the action arising from a total time derivative of the Lagrangian. 
    The $Q$ system described by \ref{action_negative_freq} is another time-dependent harmonic oscillator with unit mass and time-dependent frequency $\Omega(t)$ given by \ref{mass_sqrt_frequency}. It can be explicitly verified that: (a) The equation of motion for $Q$ namely, $\ddot Q + \Omega^2(t) Q =0$ reduces to the equation of motion for $q(t)$ obtained from the action in \ref{actiontimedependent} when we substitute $Q= \sqrt{m(t)}\, q$. (b) The time-dependent Schr\"{o}dinger equation obtained from the action in \ref{action_negative_freq} for the dynamical variable $Q$ transforms properly to the Schr\"{o}dinger equation obtained from the action in \ref{actiontimedependent} for the dynamical variable $q$. All these seem fine but note that, despite our assumption that $\omega^2(t)$ is always positive, the frequency-squared $\Omega^2$ of the new oscillator system $Q$ can be either positive or negative depending on whether $\omega^2(t)>\ddot{g}/g$ or $\omega^2(t)<\ddot{g}/g$. (See \ref{rel_to_overdamp} for a comparison 
    of this condition with that of over-damping and under-damping of the oscillator). In other words, the same physical system can be represented either as a harmonic oscillator or as an inverted harmonic oscillator depending on the dynamical variable we have chosen!

    It can be shown that $q\rightarrow Q=\sqrt{m}q$ is the unique transformation (up to a constant rescaling) that maps the oscillator system given by \ref{actiontimedependent} to a time-dependent oscillator of \textit{constant mass}. However, there is a family of such transformations parametrized by a function $f(t)$, if we also choose to change the time co-ordinate from $t\rightarrow \eta$ along with the change of variable from $q\rightarrow Q$. To understand this, let us start by introducing a new variable $Q_f=q/f(t)$, and a new time coordinate $\eta$ such that $dt=(mf^2)d\eta$. The action in \ref{actiontimedependent} then becomes,
    \begin{align}\label{Eq_q_to_Q}
    \mathcal{A}=\frac{1}{2}\int d\eta~\left[Q_f'^{2}-\tilde{\Omega} ^{2}(\eta)Q_f^{2}\right]+\frac{1}{2}\left(\frac{f'Q_f^2}{f}\right)\bigg\vert_{\eta_1}^{\eta_2}\rightarrow \frac{1}{2}\int d\eta~\left[Q_f'^{2}-\tilde{\Omega} ^{2}(\eta)Q_f^{2}\right]
    \end{align}
    where `prime' denotes derivative with respect to $\eta$ and $\eta_1$ and $\eta_2$ denotes the initial and final times, respectively. (We will see shortly that the otherwise innocuous boundary term that we dropped in \ref{Eq_q_to_Q} will introduce a phase factor of the form $\exp\{if'Q_f^2/(2f)\}$ in the relation between wave-functions of $q$ and $Q_f$ systems.) The new harmonic oscillator system $Q_f$ has unit mass and a new time-dependent frequency $\tilde{\Omega}(\eta)$ given by:
    \begin{align}\label{Eq_connection}
    \tilde{\Omega}^{2}=mf^3\left[\frac{d}{dt}\left(m\dot{f}\right)+m\omega^2f\right]
    \end{align}
    The special choice $f=1/\sqrt{m}$ corresponds to the transformation $q\rightarrow Q=\sqrt{m}q$, where we also have $\eta=t$.

    One might think that this feature is actually related to the time dependence of mass of the oscillator. However, this is not the case when one allows transformation of the time coordinate. To see this, consider the action in \ref{actiontimedependent} with $m(t)=1$. If we now  make the transformation $Q_f=q/f$ and define a new time co-ordinate by $f^2d\eta=dt$,  the frequency of the new oscillator becomes
    \begin{align}
    \tilde{\Omega}^{2}=e^{-4 h}
    \left(\omega^2+\dot{h}^2-\ddot{h}\right)
    \end{align}
    where $f=e^{-h}$, while the mass remains unity. If we choose $f(t)$ appropriately then, this $\tilde{\Omega}^{2}$ can indeed turn negative for the range of time variable such that $\ddot{h}>\omega^2+\dot{h}^2$. In other words, one can transform an oscillator with unit mass and real frequency to another one with unit mass and imaginary frequency.

    To summarize, oscillators with time-dependent mass and frequency --- described by the action in \ref{actiontimedependent} --- can be represented either with a real frequency or with an imaginary frequency corresponding to either bounded or unbounded Hamiltonians. This raises the following question: What happens to the quantum theory when the representation of the oscillator is switched from real to imaginary frequency? What happens, for example, to the discrete energy  eigenstates of the normal oscillator? Further how to address the back-reaction of this system on another semi-classical degree of freedom, when the two systems are coupled?
    The rest of the paper will be devoted to investigating these issues.
     
     The plan of the rest of the paper is as follows: In \ref{Quan_dyn} we discuss the quantum dynamics of a time dependent harmonic oscillator as well as its inverted counterpart in both Schr\"{o}dinger and Heisenberg picture. This explicitly demonstrates the subtle points involved with the inversion of the harmonic oscillator by redefinitions of variables. Following which in \ref{BkReac_overview_intuitive} we have investigated the problem of back-reaction in the context of normal as well as inverted oscillator. In particular, we show that it is possible to obtain a consistent prescription for back-reaction  starting from the principle of energy conservation. Using this prescription we demonstrate that the inversion of the oscillator leaves the physics unchanged and clarify the subtleties involved in that result.  Several detailed computations as well as implications (and caveats) as regards the quantum field theory have been presented in the appendices.

	\section{Quantum dynamics in Schr\"{o}dinger and Heisenberg pictures}\label{Quan_dyn}
	
	\subsection{Quantum theory in the Schr\"{o}dinger picture}
	
	Developing the quantum theory of a harmonic oscillator with time-dependent mass $m(t)$ and squared frequency $\omega ^{2}(t)$ is straightforward in the Schr\"{o}dinger picture. The Schr\"{o}dinger equation arising from the Lagrangian in \ref{actiontimedependent} is given by
	\begin{equation}
	\label{se1}
	i\partial_t\psi(t,q)=-\frac{1}{2m(t)}\partial_{q}^2\psi(t,q)+\frac{m(t)\omega^{2}(t)}{2}\psi(t,q)
	\end{equation} 
	Given the wave function $\psi(t_1,q)$ at an initial time $t=t_1$, this equation can be integrated to obtain the wave function $\psi(t,q)$ at any later time $t>t_1$. We will be particularly interested in the situations in which $m(t_1)\equiv m_1 >0$ and $\omega^2(t_1) \equiv \omega_1^2>0$ at the initial time $t=t_1$ and the initial state of the quantum system is the ground state of the harmonic oscillator with mass $m_1$ and the real frequency $\omega_1$. 
	
	In fact, it is fairly straightforward to construct a solution to \ref{se1} using a Gaussian
	ansatz $\psi(t,q)=\mathcal{N}(t)\exp[im\dot{\xi}q^2/(2\xi)]$, where $\xi$ satisfies the following linear differential equation \cite{mahajan2008particle},
	\begin{align}\label{eom_for_q}
	\ddot{\xi}+\frac{\dot{m}}{m}\dot{\xi}+\omega^2\xi=0
	\end{align}  
	Further, the initial condition $\dot{\xi}(t_1)=i\omega(t_1)\xi(t_1)$ and Wronskian condition $im(\xi\dot{\xi}^*-\xi^*\dot{\xi})=2\omega(t_1)$ are also imposed on the solution $\xi(t)$. (See, \ref{eom_for_xi} of \ref{solving_se1}; a brief description of the above approach  is included in this appendix for the sake of completeness.) 
	
	The Gaussian form of the wave function is preserved by the time evolution even when $\omega^2(t)$ flips sign and becomes negative at some later time. So, irrespective of whether the Hamiltonian of the system is bounded from below or not, we have well-defined, normalized, Gaussian solutions to the Schr\"{o}dinger equation describing the system. This solution starts from a ground state at $t=t_1$ and evolves as a spreading Gaussian. We emphasize this because one might think that the quantum theory of a system with unbounded Hamiltonian is pathological. There is no pathology in the above solution to the Schr\"{o}dinger equation at any finite time \textit{even when the frequency becomes imaginary}. 
	
	At any later time $t>t_1$ when $m(t)\equiv m_t>0$ and $\omega^2(t) \equiv \omega^2_t>0$, one can expand the wave function $\psi(t,q)$ in terms of the \textit{instantaneous} energy eigenfunctions, $\phi_n (q; m_t,\omega_t)$, of a harmonic oscillator with mass $m_t$ and real frequency $\omega_t$ as:
	\begin{equation}
	\psi(t,q) = \sum_{n=0}^\infty c_n \phi_n (q; m_t,\omega_t) \exp(-itE_n); \qquad E_n =\omega_t(n+1/2)
	\end{equation} 
	If the evolution freezes at time $t$, then one could have thought of the system remaining as a harmonic oscillator with mass $m_t$ and frequency $\omega_t$ at all later times.  In that case, $P(n)=|c_n|^2$ would represent the probability to find the system in the $n$th excited state of this oscillator, at all times later than $t$,  given that it started in the ground state\footnote{In the context of QFT in a homogeneous external background, one can decompose the field into a bunch of harmonic oscillators, each labelled by a wavenumber ${\bm{k}}$. It is then usual to
	 interpret $P(n_{\bm{k}})$ as the probability for $n_{\bm{k}}$ particles with wavenumber ${\bm{k}}$ to have been produced; see \ref{Appendix_A}. While this interpretation is extensively used in the literature\cite{mukhanov_winitzki_2007,popov1972,PhysRevLett.21.562,PhysRevD.2.1191,mahajan2008particle,
		calzetta2008nonequilibrium,birrell1982quantum,Dabrowski:2016tsx,Dabrowski:2014ica}, the notion of particle number in an interacting situation is ambiguous and non-covariant. One needs to invoke additional, somewhat ad hoc, criteria to use this interpretation. We will not enter into this discussion in this paper except for the comments in \ref{Appendix_A}.  We adopt this attitude because the issues we are really interested in are not related to the standard, well-known, ambiguity of particle number in a time-dependent background.} at time $t=t_1$,  due to the action of some external source which is encoded through the time dependence of $m(t)$ and $\omega(t)$. 
	
	We cannot do any of these if $\omega^2(t)<0$ at the time when we are interested in studying the state of the system. There is no natural definition of an instantaneous harmonic oscillator with \textit{real} frequency or its energy eigenstates labelled by an integer $n$. (This fact, e.g., was  crucial to count the number of particles in the QFT context,)   So, the difficulty is \textit{not} in describing the evolution of quantum system using the Schr\"{o}dinger equation when the oscillator becomes inverted. That poses \textit{no} problem and we do have a nice, normalizable Gaussian wave function describing the system at all finite times even when $\omega^2(t)$ flips sign. What fails is the interpretation of this wave function in terms of a set of natural eigenfunctions labelled by integers. 
	
	When the same system is described in the two representations --- one involving normal oscillators and the other involving inverted oscillators --- similar issues arise in the case of inverted oscillator representation. It is straightforward to show how the general transformations of the form $Q_f=q/f$ modifies the wave function viz., how the wave function $\psi(q,t)$ for $q$ system is connected to the wave function $\Psi(Q_f,\eta)$ for the $Q_f$ system.  It can be shown that 
	\begin{align}\label{Eq_psi_phi}
	\psi(t,q)=\frac{1}{\sqrt{f}}\exp\left(im\frac{\dot{f}}{f}q^{2}\right)\Psi\left[\eta(t),Q_f=q/f(t)\right]
	\end{align}
	In addition to substituting for $Q$ in terms of $q$, we have to add a phase factor arising from the total time derivative in the Lagrangian. Obviously, the Gaussian wave-functions in either representations transforms to another Gaussian wave function with an extra phase in the second representation.
	More generally, 
	the path integral kernels which can be used to evolve any wave function from an initial to final state are related to each other in the two representations by (see \ref{kernel_and_wavefn})
	\begin{align}
	\label{K_q_to_Q}
	K_{q}(q_2,t_2;q_1,t_1)=\frac{1}{\sqrt{f(t_1)f(t_2)}}\exp\left(im(t_2)\frac{\dot{f}(t_2)}{f(t_2)}q_2^{2}-im(t_1)\frac{\dot{f}(t_1)}{f(t_1)}q_1^{2}\right)K_{Q}(Q_{f1},\eta_2;Q_{f1},\eta_1)
	\end{align}
	These kernels are well-defined and exhibit no pathologies in either representations. So one can evolve any sensible initial wave function, in either representation to arrive at a wave function at a later time. The quantum theory in Schr\"{o}dinger picture is alive, well and non-pathological even when $\omega^2(t)$ flips sign and the oscillator gets inverted. As we pointed out before, the real difficulty is not in describing the quantum evolution but in introducing discrete energy eigenstates, labelled by an integer,  when the oscillator becomes inverted.
	
	The fact that quantum evolution is well-defined suggests there should exist a way of describing both normal and inverted oscillators in the Heisenberg picture as well. We  will next study the same system in the Heisenberg picture which offers further insights.
	
	\subsection{Quantum theory in the Heisenberg picture}\label{sqrt_mass_f}
	
	In \ref{inverted_oscillator}, we found that we can transform a time-dependent oscillator of frequency $\omega(t)$ and mass $m(t)$ to another time-dependent oscillator of unit mass and a new time-dependent frequency $\Omega(\eta)$ given by \ref{Eq_connection}, where the new time co-ordinate $\eta$ is defined as $mf^2d\eta=dt$. We saw that, even when $\omega^2$ is always positive, for an arbitrary choice of $f(t)$, the frequency-squared for the $Q_f$ system $\Omega^2$ can be negative in certain intervals of time. Therefore, to study the quantum dynamics of the oscillator system $Q_f$ we need a formalism in which a time-dependent harmonic oscillator of positive as well as negative frequency-squared can be dealt in a unified fashion. In this section, we will briefly motivate such a formalism.
	
	For any time-dependent harmonic oscillator, irrespective of whether the frequency $\omega$ is real or imaginary (or equivalently, whether $\omega^2$ is positive or negative) we can introduce the following operators,
	\begin{align}\label{defineaminus}
	a_{-}&=\sqrt{\frac{m\omega}{2}}\hat{q}+i\frac{\hat{p}}{\sqrt{2m\omega}}
	\\\label{defineaplus}
	a_{+}&=\sqrt{\frac{m\omega}{2}}\hat{q}-i\frac{\hat{p}}{\sqrt{2m\omega}}
	\end{align}
	For real $\omega$ it turns out that, $a_{-}^{\dagger}=a_{+}$ and $a_{+}^{\dagger}=a_{-}$ as they should, as $x$ and $p$ are hermitian operators. On the other hand, for imaginary $\omega$, i.e., for inverted harmonic oscillator it turns out that, $\omega =i\omega _{r}$, where $\omega _{r}$ is a real quantity. Hence we obtain,
	\begin{align}
	a_{-}&=\sqrt{i}\left[\sqrt{\frac{m\omega_{r}}{2}}\hat{q}+\frac{\hat{p}}{\sqrt{2m\omega _{r}}}\right]
	\\
	a_{+}&=\sqrt{i}\left[\sqrt{\frac{m\omega_{r}}{2}}\hat{q}-\frac{\hat{p}}{\sqrt{2m\omega_{r}}}\right]
	\end{align}
	From which one can immediately argue that 
	\begin{align}
	a_{-}^{\dagger}=ia_{-};\qquad a_{+}^{\dagger}=ia_{+}
	\end{align}
	Note that, irrespective of the nature of  $\omega$ the commutation relation $[a_{-},a_{+}]=1$, holds. The Hamiltonian on the other hand becomes, 
	\begin{align}
	H=\left(a_{+}a_{-}+\frac{1}{2}\right)\omega
	\label{Hdef}
	\end{align}
	In the case of inverted oscillator, with $\omega \rightarrow i\omega _{r}$, the form of Hamiltonian looks as though it is anti-Hermitian. However, we can easily verify that the Hamiltonian is indeed Hermitian ($H^{\dagger}=H$) as it should be: 
	\begin{align}
	H^{\dagger}&=-i\omega _{r}\left(a_{-}^{\dagger}a_{+}^{\dagger}+\frac{1}{2} \right)
	\nonumber
	\\
	&=-i\omega _{r}\left(-a_{-}a_{+}+\frac{1}{2} \right)
	=-i\omega _{r}\left(-a_{+}a_{-}-1+\frac{1}{2} \right)
	=H
	\end{align}
	Also note that,
	\begin{align}
	[H,a_{-}]=-i\omega _{r}a_{-};\qquad [H,a_{+}]=i\omega _{r}a_{+}
	\end{align}
	Recall that for a standard time-dependent oscillator with $\omega^2>0$ for all times, the time evolution of the creation and annihilation operators are usually described by a time-dependent Bogoliubov transformation
	\begin{align}
	a(t)=\mathcal{A}(t)a_1+\mathcal{B}(t)^*a_1^{\dagger}&&a^{\dagger}(t)=\mathcal{A}^*(t)a_1^{\dagger}+\mathcal{B}(t)a_1;\qquad\textrm{(For $\omega^2(t)>0$)}
	\end{align}
	where, $a_1\equiv a(t_1)$. The time-dependent Bogoliubov coefficients $\mathcal{A}$ and $\mathcal{B}$ are given by
	\begin{align}\label{A_and_B}
	\mathcal{A}(t)=\frac{1}{2}\sqrt{\frac{m\omega}{\omega(t_1)}}\left(\xi^*+\frac{i}{\omega}\dot{\xi}^*\right)&&
	\mathcal{B}(t)=\frac{1}{2}\sqrt{\frac{m\omega}{\omega(t_1)}}\left(\xi^*-\frac{i}{\omega}\dot{\xi}^*\right)
	\end{align}
	where, $\xi$ is a solution of \ref{eom_for_q} with the initial condition $\dot{\xi}(t_1)/\xi(t_1)=i\omega(t_1)$ and satisfies the Wronskian condition $im(\xi\dot{\xi}^*-\xi^*\dot{\xi})=2\omega(t_1)$. It is easy to verify that $\mathcal{A}\mathcal{A}^*-\mathcal{B}\mathcal{B}^*=1$. 
	
	When the sign of $\omega^2(t)$ is indefinite and, say, changes during the evolution, the relevant operators are $a_{-}$ and $a_{+}$, which were introduced in \ref{defineaminus} and \ref{defineaplus}. (These are, of course, not the conjugates of each other if $\omega$ is imaginary.) The time evolution of these operators can be described by a generalization of the standard time-dependent Bogoliubov(like) transformation, given by:
	\begin{align}
	a_{-}(t)=\mathcal{A}(t)a_1+\bar{\mathcal{B}}(t)a_1^{\dagger}&&a_{+}(t)=\bar{\mathcal{A}}(t)a_1^{\dagger}+\mathcal{B}(t)a_1
	\end{align}
	where, we have taken the initial time $t_1$ such that $\omega^2(t_1)>0$. The time-dependent coefficients are then by given by
	\begin{align}\label{bogolforinver1}
	\mathcal{A}(t)&=\frac{1}{2}\sqrt{\frac{m\omega}{\omega(t_1)}}\left(\xi^*+\frac{i}{\omega}\dot{\xi}^*\right)&&
	\mathcal{B}(t)=\frac{1}{2}\sqrt{\frac{m\omega}{\omega(t_1)}}\left(\xi^*-\frac{i}{\omega}\dot{\xi}^*\right)\\\label{bogolforinver2}
	\bar{\mathcal{A}}(t)&=\frac{1}{2}\sqrt{\frac{m\omega}{\omega(t_1)}}\left(\xi-\frac{i}{\omega}\dot{\xi}\right)&&
	\bar{\mathcal{B}}(t)=\frac{1}{2}\sqrt{\frac{m\omega}{\omega(t_1)}}\left(\xi+\frac{i}{\omega}\dot{\xi}\right)
	\end{align}
	Note that, in general $\bar{\mathcal{A}}\neq\mathcal{A}^*$ and $\bar{\mathcal{B}}\neq\mathcal{B}^*$ while, for $\omega^2(t)>0$, $\bar{\mathcal{A}}=\mathcal{A}^*$ and $\bar{\mathcal{B}}=\mathcal{B}^*$. From \ref{bogolforinver1} and \ref{bogolforinver2} it also follows that $\mathcal{A}\bar{\mathcal{A}}-\mathcal{B}\bar{\mathcal{B}}=1$. When $\omega^2(t)>0$, it is well-known that the 
	expectation value of the number operator\footnote{This will correspond to average number of `particles' or `excitations' in the context of, say, condensed matter physics. We will use this terminology with the understanding that it is just the expectation of value of the number operator without any additional attributes associated with a 'particle'. As we said before the particle concept is non-covariant and inherently ambiguous.}
	 $n(t)\equiv\braket{\textrm{in}|a_-^{\dagger}(t)a_-(t)|\textrm{in}}$ in the state $\ket{\textrm{in}}$ annihilated by $a_{1}$ is given by
	\begin{align}
	n(t)=|\mathcal{B}(t)|^2
	\end{align}  
	Since, the notion of instantaneous ground state is not defined at times when $\omega^2(t)<0$, one cannot define an average particle number at those times. With the help of operators $a_-$ and $a_+$, we will now introduce a natural generalization of the notion of particle numbers at times when $\omega^2(t)<0$. To do this, note that, in the normal case we have the relation $n=(\braket{H}/\omega(t))-(1/2)$. Since $H$ remains well-defined even for inverted oscillator, we will define:
	a quantity $n(t)$ as:
	\begin{align}\label{newnumber}
	n(t)\equiv\frac{\braket{H}}{\omega(t)}-\frac{1}{2}=
	\braket{in|a_+a_-|in}
	\end{align}
	where we have used \ref{Hdef}.
	It can be easily verified that $n(t)=\mathcal{B}(t)\bar{\mathcal{B}}(t)$. Note that $n(t)$ is in general a complex number with real and imaginary parts given by
	\begin{align}
	\textrm{Im}[n(t)]=-\frac{\braket{\textrm{in}|H(t)|\textrm{in}}}{|\omega(t)|^2}\textrm{Im}\left[\omega(t)\right];\quad
	\textrm{Re}[n(t)]=\frac{\braket{\textrm{in}|H(t)|\textrm{in}}}{|\omega(t)|^2}\textrm{Re}\left[\omega(t)\right]-\frac{1}{2}
	\end{align}
	In the case of real frequency, the imaginary part of particle number would be vanishing. On the other hand, when frequency is purely imaginary $\textrm{Im}[n(t)]$ contains all the information with $\textrm{Re}[n(t)]=-1/2$!  This may have interesting applications as far as the negative frequency harmonic oscillator is considered and provides a unified definition of particle number in the context of normal and inverted oscillators. 
	
	Given the above results, it is interesting to ask the following questions: What happens to a physical system when the function $\omega^2(t)$ crosses zero from above; that is, what happens if $\omega^2(t)$ flips sign at some instant $t=t_0$ and turns negative?
	It turns out that one can obtain reasonably general results for the behaviour of the system when this happens. 
	 Let us first analyse a class of harmonic oscillators with unit mass and a time-dependent frequency $\omega^2(t)$ which has the following behaviour near  $t_0$ at which it changes sign:
	\begin{align}
	\omega^2(t)=-\kappa(t-t_0)+\mathcal{O}\left((t-t_0)^2\right)
	\end{align}
	where, $\kappa$ is a positive constant. Hence, a solution $\xi$ of the equation $\ddot{q}+\omega^2(t)q=0$, which behaves as a positive frequency mode near some initial time $t_1$, will have the following behaviour near $t=t_0$:
	\begin{align}\label{linear}
	\xi(t)\approx N\left[\textrm{Ai}\left(\kappa^{1/3}(t-t_0)\right)+\mathcal{R}~\textrm{Bi}\left(\kappa^{1/3}(t-t_0)\right)\right]
	\end{align}
	where, $\textrm{Ai}$ and $\textrm{Bi}$ are Airy functions, $\mathcal{R}$ is a complex constant which can be determined from the initial condition at $t=t_1$ and $N=\sqrt{-\pi/(2\kappa^{1/3}\textrm{Im}[\mathcal{R}])}$ is an overall normalization. The particle number $n$ near $t=t_0$ can then be written as:
	\begin{align}
	n(t)=\frac{|(1-\sqrt{3}\mathcal{R})|^2\kappa^{2/3}|N|^2}{3^{5/3}\Gamma(1/3)^2\left[-\kappa(t-t_0)\right]^{1/2}}-\frac{1}{2}+\frac{|(1+\sqrt{3}\mathcal{R})|^2|N|^2\left[-\kappa(t-t_0)\right]^{1/2}}{3^{7/3}\Gamma(2/3)^2}+\mathcal{O}\left[(t-t_0)^{3/2}\right]
	\end{align}
	Clearly, $n(t)$ diverges as $1/\omega$ as we approach $t=t_0$ from $t<t_0$. As soon as $t>t_0$, the `particle number' $n(t)$ becomes complex with $\textrm{Re}[n]=-1/2$ and, in the neighbourhood of $t_0$, $\textrm{Im}[n]\propto 1/\omega$. 
	
	We can, in fact, demonstrate  that this feature holds in more general situations in which $\omega^2$ approaches zero in an arbitrarily smooth manner. To do this,
	let us consider a time-dependent harmonic oscillator with  $\omega^2(t)$ approaching zero smoothly as $t\rightarrow t_0$. To study particle number $n$ in this scenario, it is convenient to work with the variable $z=\mathcal{B}^*/\mathcal{A}^*e^{-2i\rho}$, where $\dot{\rho}=\omega$. From the definition of $\mathcal{B}$ and $\mathcal{A}$, one can show that $z$ satisfies the following differential equation (see \ref{derivationforz}) \cite{mahajan2008particle}:
	\begin{align}\label{eom_for_z}
	\dot{z}+2i\omega z+\frac{1}{2}\left(\frac{\dot{\omega}}{\omega}+\frac{\dot{m}}{m}\right)(z^2-1)=0
	\end{align}
	The particle number $n$ and $z$ are related as\footnote{In \cite{mahajan2008particle}, $z$ was introduced for the analysis of particle production in an external background. It is also related to the classicality of the quantum field and used extensively in the context of cosmology to understand the transition to classicality \cite{Singh:2013pxf,Singh:2013bsa,Singh:2016wmd,Das:2013qwa,Lochan:2014dca,
		Sharma:2017ivh}.} 
	\begin{equation}
	n=\frac{|z|^2}{1-|z|^2}                         
	\end{equation}  
	 Now, near $t=t_0$, \ref{eom_for_z} can be approximated  as
	\begin{align}\label{approx_eom}
	\dot{z}+\frac{1}{2}\left(\frac{\dot{\omega}}{\omega}+\frac{\dot{m}}{m}\right)(z^2-1)\approx 0
	\end{align}
	The general solution for this equation can be written as
	\begin{align}\label{soln}
	z\approx \frac{ \mathcal{C}m\omega-1}{\mathcal{C}m\omega+1}
	\end{align}
	where, $\mathcal{C}=\mathcal{C}_R+i\mathcal{C}_I$ is a complex constant. We can easily see that as $\omega\rightarrow 0$, we have $|z|\rightarrow 1$. The expression for particle number then becomes
	\begin{align}
	n(t)\approx\frac{1}{4\mathcal{C}_Rm\omega}-\frac{1}{2}+\frac{|\mathcal{C}|^2m\omega}{4\mathcal{C}_R}
	\end{align}
	Hence, we see that the particle number diverges as $1/\omega$. 
	
	On the other hand, from the relation $\braket{H}=\omega(n+1/2)$, we see that the expectation of Hamiltonian at $t=t_0$ is finite and is given by:
	\begin{equation}
	\braket{H(t_0)}=\frac{1}{(4m\mathcal{C}_R)}                                   
	\end{equation} 
	If we analytically continue the expression of $n$ for imaginary frequencies (equivalently negative $\omega^2$), we can also see that as $t\rightarrow t_0^{+}$, the frequency becomes $\omega\rightarrow i|\omega|$ and the real part of $n$ becomes $n_R\rightarrow-1/2$. The imaginary part $n_I$ on the other hand, is given by $n_I\rightarrow-\{\braket{H(t_0)}/|\omega|\}$.

	We conclude this section with a comment on a popular misconception. The transformation from normal to inverted oscillator can be thought of as a redefinition of dynamical variables (corresponding to a field redefinition in QFT). It is generally assumed that such redefinitions do not change ``physics''. While this is true at a formal level, the interpretation of the system in terms of the original dynamical variable and the redefined dynamical can be very different. (A classic example is in the case spontaneous symmetry breaking in which the two fields, related by redefinition, corresponds to two different vacua and one is certainly more physical than the other.) The discussion in this section shows clearly that the quantum theory takes a very different flavour when we switch from a normal oscillator to an inverted oscillator. This is, of course, expected  since several standard features of normal oscillator (e.g., equally spaced, discrete energy levels) disappear when we introduce such a field redefinition. On the other hand, some other physical features, like the back-reaction of this system on another system to which it is coupled, should not lead to new physical phenomena under such field redefinitions. At this stage, it is far from clear how this comes about and this issue requires detailed investigation. We now take up several aspects of this problem.
	
	\section{Quantum back-reaction on the classical source} \label{BkReac_overview_intuitive}
	
	From the study of the time-dependent oscillator and inverted oscillator in the Heisenberg picture and in the Schr\"{o}dinger picture, we arrive at the conclusion that while the quantum theory remains well-defined, its interpretation in terms of discrete set of energy eigenstates becomes ambiguous since it exists only in the normal representation.\footnote{In the context of QFT, when one tries to use these energy eigenstates to define particle number, this introduces a \textit{new} level of ambiguity. It is, of course, well-known that the notion of particles in an external field is fraught with conceptual issues. Usually such issues are related to ambiguity in the notion of particles (e.g., the definition of positive-frequency mode may not be unique). But in this case, the natural definition of particle \textit{itself} ceases to exist in one representation though it exists in another representation.} While this might appear rather disturbing at first sight, what is really relevant are the physical effects of such an interpretation. If the oscillator is excited by the action of an external source in one representation, the energy for the excitation has to come from the external source which will reflect in a suitable form of back-reaction on the source. 
	In the second representation with an inverted oscillator, the quantum evolution remains sensible but there is no interpretation in terms of particle excitations. This raises the question as to how we can describe the back-reaction when we are using the inverted oscillator representation. Since the two representations describe the same physical system, we would expect the back-reaction to remain the same so that all physical effects remain representation independent. We will now see how to formulate the back-reaction equations in such a way that this desirable feature is indeed realized. 
	
	\subsection{Back-reaction from energy conservation}\label{BR_from_EC}
	
	The idea of classical (i.e., c-number) back-reaction on a source due to quantum effects is a conceptually difficult issue and, in general, is a somewhat ill-posed problem. We will therefore try to approach it from solid physical considerations. One such approach is to determine the back-reaction from demanding that the energy needed for the quantum excitation (which will translate to both particle production and change in vacuum polarization in the QFT language) must be supplied by the classical source. Let us first see what this leads to.
	
	In order to do this, consider the following Lagrangian
	describing two degrees of freedom, denoted by $C(t)$ and $q(t)$, where, eventually we want to think of $q(t)$ as quantum mechanical and $C(t)$ as classical:
	\begin{align}
	\mathcal{L}(C,q)=\frac{\mathcal{M}}{2}\dot{C}^2-\mathcal{M} V(C)+\frac{m(C)}{2}\dot{q}^2-U(C,q)
	\end{align}
	(The reason for scaling the potential for $C$ by $\mathcal{M}$ will become clear soon).
	The Hamiltonian of the system is then given by 
	\begin{align}\label{hamiltonianforcq}
	\mathcal{H}= \frac{P^2}{2\mathcal{M}}+\mathcal{M} V(C)+\frac{p^2}{2m(C)}+U(C,q)
	\end{align}
	where $P=\partial_{\dot{C}}\mathcal{L}$ and $p=\partial_{\dot{q}}\mathcal{L}$ are the canonical momenta corresponding to $C$ and $q$, respectively. Given the Hamiltonian, one can define in a straight forward manner the following two levels of dynamics: 
	
	(a) The full quantum dynamics, in which we treat both the degrees of freedom $C$ and $q$ as quantum mechanical. Such a situation can be described, say, by a wave function $\Psi(t,q,C)$ determined by the Schr\"{o}dinger equation based on the full Hamiltonian. 
	
	(b) The classical dynamics in which we treat both the degrees of freedom $C$ and $q$ as fully classical.  The deterministic evolution is then governed by standard Euler-Lagrange or Hamilton's equations solved by two functions $C(t)$ and $q(t)$. There is a general procedure to obtain the classical description from the quantum mechanical one in the limit of $\hbar\to0$ if the quantum state can be well approximated by a form $\Psi=R\exp( iS)$. 
	
	It is also possible to consider a \textit{third kind} of description in which $C$ is treated as classical and $q$ is treated as quantum mechanical.
	Such a description is also conceptually simple \textit{if we ignore the back-reaction} of $q$ on $C$. Then dynamics of the system is described by the equations:
	\begin{equation}
	M(\ddot C+V'(C))\approx0; \qquad i\hbar\partial_t\psi=h\psi                                                           
	\end{equation} 
	The first is
	the classical equation $M(\ddot C+V'(C))\approx0$ unaffected by the quantum variable. The second is a Schr\"{o}dinger equation\footnote{This description is analogous, say, to QFT in FRW spacetime. The $C$ corresponds to $a(t)$, the equations $M(\ddot C+V'(C))\approx0$ correspond to the Einstein's equations determining $a(t)$ and $h(p,q,t)$ will correspond to the Hamiltonian of a scalar field in a given FRW background.} with the time-dependent Hamiltonian $h(p,q,t)=\{p^2/2m(C(t))\}+U(C(t),q)$.  This limit, however,  cannot be obtained from the full quantum theory by $\hbar\to0$ approach. We need another small parameter in the problem to define this approximation, closely related to the Born-Oppenheimer approximation in molecular dynamics \cite{born1924quantum}. (In molecular dynamics this small parameter is the ratio of masses of electron and nuclei and allows such a description. In our case, we will use $1/\mathcal{M}$ in the Lagrangian as the relevant small parameter and obtain an  analogous description  \textit{without} back-reaction \cite{Kiefer:1991ef,hartle1981,rubekov1979,Padmanabhan:1988ur,Padmanabhan:1990fn}).

	What we are interested in is a \textit{fourth kind} of description which can be thought of as the next order approximation to the above one. In this, we want to take into account the back-reaction of the quantum dynamics on the classical background by some effective c-number description. This is more non-trivial and, for the purpose of this paper, we will adopt the following procedure.
	It has been demonstrated in \cite{singh1989notes} that we can construct solutions of the Schr\"{o}dinger equation of the $C-q$ system such that as $1/\mathcal{M}\rightarrow 0$, the dynamics of $C$ system is effectively classical while $q$ remains quantum mechanical. In light of this, let us consider a quantum state of the $C-q$ system $\ket{\tilde{\psi},\mathcal{M}}$ such that there exist another state vector $\ket{\psi}$ in the Hilbert space of the system $q$ and a real function $C(t)$ such that $\braket{\tilde{\psi},\mathcal{M}|\mathcal{H}(\hat{C},\hat{P},\hat{q},\hat{p})|\tilde{\psi},\mathcal{M}}=\braket{\psi|\mathcal{H}(C(t),P(\dot{C}(t)),\hat{q},\hat{p})|\psi}+\mathcal{O}(1/\mathcal{M})$, where $\ket{\psi}$ satisfies the following Schr\"{o}dinger equation. 
	\begin{align}
	\left[\frac{\hat{p}^2}{2m(C)}+U(C(t),\hat{q})\right]\ket{\psi}=i\partial_{t}\ket{\psi}
	\end{align}
	There is strong evidence for the existence of such states, even though there are still some unresolved issues (for example see \cite{singh1989notes}). In this case, the expectation value of $\mathcal{\hat{H}}$ with respect to $\ket{\tilde{\psi},\mathcal{M}}$ in the limit of large $\mathcal{M}$ is given by
	\begin{align}\label{tot_E}
	E(t)=\braket{\tilde{\psi},\mathcal{M}|\mathcal{H}|\tilde{\psi},\mathcal{M}}\approx\frac{\mathcal{M}\dot{C}^2}{2}+\mathcal{M} V(C)+\left[\frac{\bra{\psi}\hat{p}^2\ket{\psi}}{2m(C)}+\bra{\psi}U(C,\hat{q})\ket{\psi}\right]
	\end{align}
	That is, the mean energy of these quasi-classical states can be approximated as the sum of (a) the classical expression for the energy of $C$-mode
	(the first two terms in the right hand side) and (b) the expectation value of the quantum Hamiltonian in this classical background. It is intuitively obvious that if the relevant limit --- viz. the one in which we can think of $C$ as classical and $q$ and quantum --- exists then it should occur for the quasi classical states which satisfy this criterion.
	The conservation of the total average energy $E(t)$ then implies that $dE/dt=0$. This can be rewritten, using  \ref{tot_E}, as
	\begin{align}\label{conserve_E}
	\dot{C}\left(\mathcal{M}\ddot{C}+\partial_CV(C)\right)+\frac{d}{dt}\braket{\psi|\mathcal{H}_q(\hat{q},\hat{p};C)|\psi}=0
	\end{align} 
	where, $\mathcal{H}_q(\hat{q},\hat{p};C)\equiv\hat{p}^2/(2m(C))+U(C,\hat{q})$ is the Hamiltonian for $q$-system in the classical background $C$. In the Heisenberg picture, we can express the second term of \ref{conserve_E} as $\braket{\psi|\dot{\mathcal{H}}_q(\hat{q},\hat{p};C)|\psi}$. Further, the total time derivative of the Hamiltonian $\mathcal{H}_q$ is just the partial derivative $\partial\mathcal{H}_q/\partial t$. Since the explicit time dependence of $\mathcal{H}_q$ comes through it's dependence on the classical background $C$, one can write $\partial\mathcal{H}_q/\partial t=\dot{C}\partial\mathcal{H}_q/\partial C$. Hence, \ref{conserve_E} can be simplified to
	\begin{align}\label{conserve}
	\mathcal{M}\ddot{C}+\mathcal{M}\partial_CV(C)+\left[-\frac{\partial_Cm(C)}{2m^2(C)}\braket{\psi|\hat{p}^2|\psi}+\braket{\psi|\partial_CU(C,\hat{q})|\psi}\right]=0
	\end{align}
	Now, if we assume that the initial condition at $t=t_1$ on $\ket{\tilde{\psi},\mathcal{M}}$ is such that $\ket{\psi}\big\vert_{t_1}=\ket{\textrm{in}}$, where $\ket{\textrm{in}}$ is the instantaneous vacuum of the $q$ system in the presence of a classical background $C(t)$ then, \ref{conserve} reduces to the so called `in-in' prescription for back-reaction. 
	
	We will apply these ideas to a classical `particle'  having mass $\mathcal{M}$ and moving in a potential $\mathcal{M}V(C)$, while it is coupled to the quantum system in a quadratic manner. We will take the Lagrangian for the system as,
	\begin{align}\label{Lag_quad}
	L(C,q)=\frac{1}{2}\mathcal{M}\dot{C}^{2}-\mathcal{M}V(C)+\frac{1}{2}m(C)\left(\dot{q}^{2}-\omega^{2}(C)q^{2}\right)
	\end{align}
	where $C$ is the classical degree of freedom while $q$ stands for the quantum operator. Note that the Lagrangian for the quantum part is equivalent to that of a harmonic oscillator with a variable mass $m(C)$ and variable frequency $\omega^2(C)$ in a background $C(t)$.
	The `in-in' prescription for back-reaction, given by \ref{conserve}, for this system leads to the following description.  The \textit{semi-classical} equation of motion for $C$ in the presence of the quantum operator using \ref{conserve} is given by, 
	\begin{align}\label{bk_reac}
	\mathcal{M}\ddot{C}+\mathcal{M}\partial_C V(C)-\frac{\partial_Cm(C)}{2m^2(C)}\braket{\textrm{in}|\hat{p}^2|\textrm{in}}+\frac{1}{2}\partial_C\left(m(C)\omega^2(C)\right)\braket{\textrm{in}|\hat{q}^2|\textrm{in}}=0
	\end{align}
	where the state $|\textrm{in}\rangle$ corresponds to the `in' vacuum of the $q$ system in a classical background $C(t)$ and is defined by
	\begin{align}\label{define_in}
	a_1\ket{\textrm{in}}=0
	\end{align}
	
	On the other hand, as far as the quantum degree of freedom $q$ is concerned it essentially resembles a harmonic oscillator with time-dependent mass $m(C)$ and frequency $\omega(C)$.  Therefore, if one can understand the time-dependent oscillator problem to some detail, it will certainly shed more light on the behaviour of quantum
	field in classical backgrounds. Using the expansion of $a_-(t)$ and $a_+(t)$ given in \ref{bogolforinver1} and \ref{bogolforinver2}, we can rewrite the \ref{bk_reac} as
	\begin{align}\label{conserve_quad}
	\frac{d}{dt}\left[\frac{\mathcal{M}}{2}\dot{C}^{2}+\mathcal{M}V(C)+\left(n+\frac{1}{2}\right)\omega \right]=0
	\end{align}
	This has the nice interpretation that the back-reaction of $q$ on $C$ has two contributions: (a) coming from the rate of change of instantaneous ground state energy of $q$, namely $\dot{\omega}$/2 and (b) coming from the rate of energy drain due to particle production, namely $d(n\omega)/dt$. From \ref{conserve_quad}, the total energy drain $\Delta E_{c}$ from the classical system from $t=t_1$ to $t=T$ is given by
	\begin{align}
	\Delta E_{c}^{\rm in-in}=-\left(\omega(T)n(T)+\frac{\omega(T)}{2}-\frac{\omega(t_1)}{2}\right)
	\end{align}
	The total energy drained from the classical system is therefore found to have two parts: (i) a part that is proportional to the average `$q$'-particles produced at $t=T$ and (ii) the change in instantaneous ground state energies from $t=t_1$ to $t=T$. So all these make physical sense (as a prescription for back-reaction) when $m>0$ and $\omega^2>0$. (See \ref{compare_ii_oi} for a brief comparison of energy conservation in `in-in' and `in-out' prescriptions).  We will now discuss how to use such a prescription consistently when we use the inverted oscillator representation.
	
	\subsection{Comparison of back-reaction in terms of $q$ and $Q$ systems}
	
	In \ref{inverted_oscillator}, we saw that the harmonic oscillator system $q$ given in \ref{eom_for_q} under the change of variable $Q=\sqrt{m}q$ transforms to a new harmonic oscillator with unit mass and a time-dependent frequency $\Omega^2$ given by \ref{mass_sqrt_frequency}. 
	 Therefore, it is important to know whether the  back-reaction prescription leads to a consistent picture under the transformation $Q\rightarrow \sqrt{m}q$. We will explicitly show below that the back-reaction equations obtained by working with variables $Q$ and $q$ are equivalent.  
	\par The wave function $\psi_C(q,t)$ for the `in-vacuum' of the time-dependent oscillator \ref{eom_for_q} can be solved using a Gaussian ansatz to get
	\begin{align}
	\braket{q|\textrm{in($t$)}}_{C}\equiv\psi_C(q,t)=\mathcal{N}\exp\left(\frac{i m\dot{\xi}_C(t)}{2\xi_C}q^2\right)
	\end{align}
	where, the `$C$' in the subscript denotes that the quantities are evaluated for a particular classical background $C(t)$ and $\xi_C$ is a solution of the time-dependent harmonic oscillator equation
	\begin{align}\label{eom_for_q1}
	\ddot{\xi_C}+\frac{\dot{m}}{m}\dot{\xi_C}+\omega^2\xi_C=0
	\end{align}  
	with $m(t)\equiv m(C(t))$ and $\omega(t)\equiv \omega(C(t))$ respectively (see \ref{solving_se1} for the details) \cite{mahajan2008particle}. Further, the initial condition $\dot{\xi}_C(t_1)=i\omega(t_1)\xi_C(t_1)$ and Wronskian condition $im(\xi_C\dot{\xi}_C^*-\xi_C^*\dot{\xi}_C)=2\omega(t_1)$ are also imposed. The back-reaction equation presented in \ref{bk_reac} then takes the form
	\begin{align}\label{bk_reac_quad_explicit}
	\mathcal{M}\ddot{C}+\mathcal{M}\partial_{C}V&-\frac{1}{2}\partial_Cm|\dot{\xi}_C|^2+\frac{\partial_C(m\omega^2)}{2}|\xi_C|^2=0
	\end{align}
	Our aim is to obtain the corresponding description in the $Q$-representation, compare the two and prove their equivalence when the back-reaction prescription is based on energy conservation. There are some interesting curiosities and subtleties which arises when we attempt this which we will first discuss. While this is all essentially classical mechanics, these issues do not seem to have received sufficient attention in the literature. (See \ref{classical_dyn} for a brief review of the classical dynamics of $C-Q$ system). The key new feature arises from the following fact: The $C-q$ Lagrangian has the form $L(\dot C,C,\dot q,q)$ which is completely standard. But the natural $C-Q$ Lagrangian has the form $L_2(C,\dot C,\ddot C,Q,\dot Q)$ containing $\ddot C$ albeit linearly.\footnote{In the case of a scalar field in FRW background the Lagrangian for the field only depends on $a$ if we use the standard  scalar field modes $\Phi_k$ as the dynamical variables. But when we use the Mukhanov-Sasaki variable $\vartheta_k$, the Lagrangian depends on $a''$. If we add to this Lagrangian the 
		Einstein-Hilbert Lagrangian for the minisuperspace described by $a(\eta)$ then the dynamics of the background is determined by a Lagrangian of the functional dependence $L(a,a',a'',\vartheta_k,\dot\vartheta_k)$. This is just a special case of a very general situation which we discuss below.} So we need to generalize our concepts to a Lagrangian with $\ddot C$.
	
	As one can see, this is a quantitative change in the structure of the dynamics when we want to incorporate the back-reaction. In the absence of back-reaction we are dealing with the dynamics of the $Q$ degree of freedom in the time-dependent background generated by $C(t)$. The fact that the Lagrangian depends on $\ddot C$ through  $\Omega^2(C,\dot C, \ddot C)$ is completely irrelevant because we just think of $\Omega^2$ as a given function of $t$ arising through a function $C(t)$ determined by the equation $\ddot C+V'(C)\approx0$, when we ignore back-reaction. (For example, this is what we do when we study either $\Phi_k$ or $\vartheta_k$ in a given FRW background with an $a(\eta)$ determined by Einstein's equations without back-reaction.) But when we want to study back-reaction both $C$ and $Q$ are dynamical variables linked through energy conservation, say. Then, when we vary $C$ in the Lagrangian to get its equations of motion (in order to see how $Q$ affects it) it is necessary to take into account the 
	existence of $\ddot C$ in the full Lagrangian.
	
	Let us briefly see how this can be done.
	We start with the Lagrangian in \ref{Lag_quad} which, under $q\rightarrow Q$, transforms to
	\begin{align}\label{lag_quad_new}
	L_2(C,\dot{C},\ddot{C},Q,\dot{Q})=\mathcal{M}\frac{\dot{C}^2}{2}-\mathcal{M}V(C)+\frac{1}{2}\left[\dot{Q}^2-\Omega^2(C,\dot{C},\ddot{C})Q^2\right]
	\end{align}
	where we have dropped a boundary terms, and  
	\begin{align}
	\Omega^2(C,\dot{C},\ddot{C})=\omega^2(C)-\frac{\ddot{C}\partial_Cm}{2m}+\frac{\dot{C}^2}{2}\left(\frac{(\partial_Cm)^2}{2m^2}-\frac{\partial^2_Cm}{m}\right)
	\end{align}
	In general the Euler-Lagrange equation for a Lagrangian of the form $L(\ddot x,\dot x,x)$ is given by:
	\begin{equation}
	\frac{\partial L}{\partial x}-\frac{d}{dt}\left(\frac{\partial L}{\partial \dot{x}}\right)+\frac{d^2}{dt^2}\left(\frac{\partial L}{\partial \ddot{x}}\right)=0
	\end{equation} 
	In our case, 
	it is straightforward to show that these Euler-Lagrange equation for $C$ obtained from $L_2$ is given by
	\begin{align}\label{puthya_EOM}
	\mathcal{M}\ddot{C}-\mathcal{M}\partial_CV-\frac{\partial_Cm}{2m}\dot{Q}^2+\frac{1}{2}\partial_C(\omega^2)Q^2+\frac{1}{2}\frac{(\partial_Cm)^2}{m^2}Q\dot{Q}\dot{C}+\frac{1}{2}\frac{\partial_Cm}{m}\omega^2Q^2-\frac{1}{8}\frac{Q^2\dot{C}^2(\partial_Cm)^3}{m^3}=0
	\end{align}
	Note that the equation is of second order in the dynamical variable even though the Lagrangian has a $\ddot C$. This is a well-known result and arises from the fact that the Lagrangian depends on $\ddot C$ only linearly and can be eliminated by adding a total time derivative reducing the Lagrangian to the standard form $L(\dot C,C,\dot Q,Q)$. Since the Euler derivative of a total derivative term is zero the equations of motion remain the same and is of second order.
	
	The equation of motion given by \ref{puthya_EOM} arises from a conserved energy function $E_{C,Q}$ quadratic in $\dot C$. There are several ways to obtain the energy function and the most natural one  use the fact that the Lagrangian $	L_2(C,\dot{C},\ddot{C},Q,\dot{Q})$ --- though has a $\ddot C$ --- has no explicit time dependence. The symmetry under time translation will lead to an energy function given by (see \ref{classical_dyn} for details):
	\begin{align}\label{tot_E_CQ}
	E_{C,Q}(t)\equiv\frac{1}{2}\mathcal{M}\dot{C}^{2}+\mathcal{M}V(C)+\frac{1}{2}\dot{Q}^{2}+\frac{1}{2}\omega ^{2}Q^{2}+\frac{\dot{m}^{2}}{8m^{2}}Q^{2}-\frac{\dot{m}}{2m}Q\dot{Q}
	\end{align}
	It can be directly verified that the condition $dE_{C,Q}(t)/dt=0$, along with the equation of motion for $Q$ leads to \ref{puthya_EOM}.
	
	Once we have a classical energy function for the system it is easy to state the prescription for the quantum back-reaction from energy conservation. In the limit of $Q$ being quantum mechanical while $C$ is effectively classical, from \ref{tot_E_CQ}, a natural definition for the total average energy of the $C-Q$ system is given by 
	\begin{align}
	E(t)\equiv\frac{1}{2}\mathcal{M}\dot{C}^{2}+\mathcal{M}V(C)+\frac{1}{2}\braket{\textrm{in}|\hat{P}^{2}|\textrm{in}}+\frac{1}{2}\omega ^{2}\braket{\textrm{in}|\hat{Q}^{2}|\psi}+\frac{\dot{m}^{2}}{8m^{2}}\braket{\textrm{in}|\hat{Q}^{2}|\textrm{in}}-\frac{\dot{m}}{4m}\braket{\textrm{in}|(\hat{Q}\hat{P}+\hat{P}\hat{Q})|\textrm{in}}
	\end{align}
	We have written the classical $PQ$ as $1/2(PQ+QP)$ to ensure Hermiticity.
	Then, the back-reaction equation that follows from $dE/dt=0$ can be written as  
	\begin{align}\label{puthya_EOM2}
	\mathcal{M}\ddot{C}-\mathcal{M}\partial_CV-\frac{\partial_Cm}{2m}\braket{\hat{P}^2}+\frac{1}{2}\partial_C(\omega^2)\braket{\hat{Q}^2}+\frac{1}{4}\frac{(\partial_Cm)^2\dot{C}}{m^2}\braket{\hat{Q}\hat{P}+\hat{P}\hat{Q}}+\frac{1}{2}\frac{\partial_Cm}{m}\omega^2\braket{\hat{Q}^2}-\frac{1}{8}\frac{\braket{\hat{Q}^2}\dot{C}^2(\partial_Cm)^3}{8m^3}=0
	\end{align}
	where, we have used the Heisenberg equation of motion for $Q$ to replace $\dot{\hat{P}}$ with $-\Omega^2 \hat{Q}$ and the expectation value is taken with respect to $\ket{\textrm{in}}$. 
	
	In order to compute these expectation values, in the Schr\"{o}dinger picture, we need to find the time-dependent wave function for the `in-vacuum', namely $\Psi_C(Q,t)\equiv\braket{Q|\textrm{in}(t)}$, where the `$C$' in the subscript denotes that the wave function is evaluated in a classical background configuration of $C$. 
	It can be easily found, by using \ref{Eq_psi_phi}, that $\Psi_C(Q,t)$ is given by
	\begin{align}\label{ground_Q}
	\Psi_C(Q,t)=\bar{\mathcal{N}}(t)\exp\left(\frac{i\dot{\mu}_C}{\mu_C}Q^2\right)
	\end{align}
	where, $\mu_C=\sqrt{m}\xi_C$ and $\bar{\mathcal{N}}(t)$ is the normalization factor. Having found the wave function, we can easily find the following expectation values of operators
	\begin{align}\label{useful_exp}
	\braket{\hat{Q}^2}=|\mu_C|^2&&\braket{\hat{P}^2}=|\dot{\mu}_C|^2 &&\braket{\hat{Q}\hat{P}+\hat{P}\hat{Q}}=\frac{d}{dt}|\mu_C|^2
	\end{align}
	In order to obtain the explicit form of the back-reaction equation in terms of $\mu_C$, we can use these expectation values in \ref{puthya_EOM2} 
	to get
	\begin{align}\label{puthya_EOM2_simplified}
	\mathcal{M}\ddot{C}-\mathcal{M}\partial_C V-\frac{\partial_Cm}{2m}|\dot{\mu}_C|^2+\frac{1}{2}\partial_C(\omega^2)|\mu_C|^2+\frac{1}{4}\frac{(\partial_Cm)^2\dot{C}}{m^2}\frac{d}{dt}|\mu_C|^2+\frac{1}{2}\frac{\partial_Cm}{m}\omega^2|\mu_C|^2-\frac{1}{8}\frac{|\mu_C|^2\dot{C}^2(\partial_Cm)^3}{m^3}=0
	\end{align}
	To compare this with the back-reaction equation equation in the $q$ representation, we can express \ref{puthya_EOM2_simplified} in terms of $\xi_{C}$ by making use of the fact that $\xi_{C}=\mu_{C}/\sqrt{m}$. By direct substitution, one can obtain the following useful identities.
	\begin{align}\label{identity1}
	-\frac{\partial_Cm}{2m}|\dot{\mu}_C|^2+\frac{1}{4}\frac{(\partial_Cm)^2\dot{C}}{m^2}\frac{d}{dt}|\mu_C|^2-\frac{1}{8}\frac{|\mu_C|^2\dot{C}^2(\partial_Cm)^3}{m^3}&=-\frac{1}{2}(\partial_Cm)|\dot{\xi}_C|^2\\\label{identity2}
	\frac{1}{2}\partial_C(\omega^2)|\mu_C|^2+\frac{1}{2}\frac{\partial_Cm}{m}\omega^2|\mu_C|^2&=\frac{1}{2}\partial_{C}(m\omega^2)|\xi_{C}|^2
	\end{align}
	Using \ref{identity1} and \ref{identity2} in \ref{puthya_EOM2_simplified}, we exactly reproduce the back-reaction equation in the $q$ representation namely \ref{bk_reac_quad_explicit}. This demonstrates that, even though the particle interpretations are different in the $q$ and $Q$ representations of the quantum degree of freedom, the back-reaction equations are equivalent.

	\section{Discussion}
	
	There are several physical systems which can be modelled as a harmonic oscillator  with time-dependent mass, $m(t)$,  and squared-frequency $\omega^2(t)$. By a transformation of the dynamical variable one can map an oscillator with positive mass and positive squared-frequency to another (inverted)  oscillator with positive mass but negative squared frequency. (A well-known example of this arises in the study of a scalar field in an expanding universe when we switch from the Fourier modes of the scalar field to the Fourier modes of the Mukhanov-Sasaki variable; but this is just a special case of a \textit{very} general phenomenon.) 
	
	When both mass and squared frequency are positive one can introduce the concept of particle-like expectations in a natural manner and hence study the excitation of the system by the external source. Such an approach fails when $\omega^2(t)<0$. We have explored this situation in detail in this paper by starting with a generic description (applicable to either sign of $\omega^2$) and comparing the features of both normal and inverted oscillators. It turns out that the behaviour of the system close to the time when $\omega^2(t)$ flips sign is universal and can be analysed completely. We show that while the standard description in terms of discrete, instantaneous,  energy eigenstates down in the inverted oscillator representation, the expectation value of Hamiltonian remains well-defined and the theory does not exhibit any pathologies. 
	
	The situation becomes more complicated when the harmonic oscillator is coupled to another degree of freedom $C$ --- so that the full system is described by a Lagrangian $L(C,\dot C,q,\dot q)$ --- and  we want to study the quantum back-reaction of $q$ on the classical degree of freedom $C$. The excitation of the $q$-degree of freedom, for example, will cost energy which has to supplied by the external source $C$. The energy conservation then requires the dynamics of the source to be modified by a back-reaction term in its (effective) equations of motion. After discussing the different limits of the $C-q$ system, we motivated a possible description of back-reaction based on energy conservation by using suitable expectation values for the quantum degree of freedom. We show that this works perfectly well --- and gives physically meaningful results  --- when squared-frequency is positive.
	
	When we change the representation to that of an inverted oscillator, two new features arise: (a) We no longer have an interpretation in terms of instantaneous energy eigenstates labelled by integers. (b) The Lagrangian now depends on the second time derivatives of $C$. But since both the oscillator representation and the inverted oscillator representation describe the same physical system we expect the back-reaction equation to be the same. We show how this can be achieved in our prescription based on energy conservation. We use the fact that, since the Lagrangian has no explicit time-dependence one can indeed define a conserved energy function $E$ for the full system such that $dE/dt=0$, which --- in turn --- reproduces the equations of motion. We construct the appropriate energy function and replace the contribution from the quantum degree of freedom by the relevant expectation value, just as we did in the case of the normal oscillator. We explicitly verify that this procedure leads to the same back-reaction equation in the inverted oscillator representation which was originally obtained in the case of the standard oscillator representation.
	
	At this stage it will be worthwhile to highlight the new results in this paper and their significance. These are embedded throughout the above discussion but we believe it is important to collect them together and state them briefly.

\begin{itemize}

\item The fact that one cannot define the notion of particles in a time-dependent background is well-known for decades. But this is addressed in the literature \textit{only} in the context of time-dependent {\it real} frequencies with $\omega^2>0$. One key result we have stressed throughout the paper is the fact that \textit{the same harmonic oscillator can be repeated either with real frequency or as imaginary frequency by a field redefinition}. This leads to yet another ambiguity in using the `particle' notion in external backgrounds.To the extent we know, this has not   been discussed in the literature at all in a focused fashion.

\item In fact, most of the current literature treats any imaginary  frequency oscillator (with $\omega^2<0$) as an unstable system. One main goal of this paper is to correct this misunderstanding and point out that the field redefinition can transform an oscillator with real frequency (with $\omega^2>0$) into an oscillator with imaginary frequency (with $\omega^2<0$). This conceptual realization is, by itself, quite significant and has not been noticed in the previous literature, as far as we know.

\item This fact has several further implications. To begin with, we know that the physics of the back-reaction from the quantum system should not depend on any redefinition. But how this invariance actually comes about, is far from obvious and needs to be demonstrated. We have done this explicitly. This is a significant step because we see that, even though a harmonic oscillator may appear at first sight to be unstable (with $\omega^2<0$) its back-reaction does not induce any runaway process.

\item The back-reaction has another non-trivial feature which is not addressed in the previous literature. This arises from the fact that, the  redefinition of the quantum dynamical variable introduces second time derivative of the classical dynamical variable into the action. Again, as far as we know, this fact and its implications, have been completely overlooked in the literature.\footnote{For example, in the context of inflationary physics, if one uses - as one often does - the Mukhonov-Sasaki variable (rather than the original scalar field), the action for the scalar field picks up second time derivative of the expansion factor. So, when we vary the total action of the (gravity + scalar field) system in the minisuperspace approximation, to obtain the back-reaction term, it is necessary to take into account the contributions from this second time derivative.} As we all know, one rarely deals with actions with second time derivative in conventional physics. Therefore the discussion in this paper assumes special significance where we show that, not only this feature can be handled properly but also that it is essential to do it if we have to obtain the correct back-reaction.

\item The study also clarified several aspects of field redefinitions and its consequences. As we mentioned earlier while a field redefinition cannot fundamentally change the nature of the observable results, the interpretation of the system in terms of two different fields (normal and inverted oscillator) will be very different. When the system is coupled to another semi-classically, the physical equivalence of the two representations arise in a fairly subtle and non-trivial manner. We have explicitly demonstrated this equivalence as regards the back-reaction.

\end{itemize}
	
	Finally we comment on a few more aspects of our analysis which we will discuss in detail elsewhere \cite{bkreactn}. Just as one can transform an oscillator  with positive mass and squared-frequency to an inverted  oscillator, one can also transform an oscillator with positive mass and positive squared-frequency to another oscillator with unit mass and \textit{constant} positive frequency. This is just a special case of what was studied here and is achieved by setting $\Omega^2=$ constant in \ref{Eq_connection} and hence solving for $f$. Such a transformation of a time-dependent oscillator to a constant frequency oscillator has several interesting consequences of which we mention two --- (a) The constant frequency oscillator has a conserved energy which can be re-expressed in terms of the original variables, thereby providing a conserved quantity for the original, time-dependent, oscillator. This conserved quantity happens to be what is known in the literature as Ermakov-Lewis invariant. This analysis demystifies the origin of such a conserved quantity for a time-dependent oscillator \cite{tpel}. (b) There is natural definition of positive frequency modes --- and the associated vacuum --- for a constant frequency oscillator while it is not easy to define the same for the time-dependent oscillator. By transforming the latter to the former, one can define a natural vacuum state for any time-dependent oscillator. For example, such an exercise, carried out for de Sitter spacetime, naturally picks up the Bunch-Davies vacuum and, for an arbitrary FRW model, it leads to a natural generalization of the Bunch-Davies vacuum \cite{bkreactn}.

	\subsection*{Acknowledgements}
	Research of S.C. is supported by the SERB-NPDF grant (PDF/2016/001589) from SERB, Government of India.
	T.P's research is partially supported by the J.C.Bose Research
	Grant of DST, India. We thank the referee for useful comments.
	
	\section*{Appendices}
	
	\appendix 
	\labelformat{section}{Appendix #1} 
	\labelformat{subsection}{Appendix #1} 
	
	\section{Connection with QFT and caveats}\label{Appendix_A}

	One physical system extensively discussed in the literature, which can be mapped to a system of uncoupled harmonic oscillators   with time dependent mass and frequency, is provided by a scalar field in an expanding background. Such a system, however, has infinite number of degrees of freedom which introduces several extra complications which are \textit{not} present in a \textit{finite} dimensional system discussed in the main body of the text. In this Appendix, we will describe the context in which our analysis can be related to that of a scalar field in an expanding background and will discuss briefly some of the caveats in using such a correspondence. 
	  
At the level of \textit{classical} field theory in an expanding background, the mathematics proceeds smoothly.	  
  Consider a quantum scalar field $\Phi(t,\bm{x})$ in an external gravitational field described by the metric $g_{\mu\nu}$. We take the action to be 
    \begin{align}
    \mathcal{A}=\int d^{4}x\sqrt{-g}\left[-\frac{1}{2}g^{\mu \nu}\partial _{\mu}\Phi \partial _{\nu}\Phi
    -\frac{1}{2}M^{2}\Phi^{2}\right]~.
    \end{align}
   In the case of a spatially flat FRW universe, the metric is given by
    $ds^2=g_{\mu\nu}dx^{\mu}dx^{\nu}=-dt^2+a(t)(d\mathbf{x})^2$, which is spatially homogeneous. Taking advantage of this special feature, one can  introduce the spatial Fourier transform $\Phi_{\mathbf{k}}(t)$ of the scalar field $\Phi(t,\bm{x})$ as follows:
    \begin{align}
    \Phi(t,\mathbf{x})=\int\frac{d^3k}{(2\pi)^3}e^{i\mathbf{k}.\mathbf{x}}\Phi_{\mathbf{k}}(t)~.
    \end{align}
    This allows us to reduce the above problem of a quantum field in FRW background to that of a bunch of harmonic oscillators, each labelled by the wave vector ${\mathbf{k}}$. The Lagrangian for each oscillator  is given by 
    \begin{align}\label{actionforscalarinfrw}
    L=\left[\frac{a^{3}}{2}|\dot{\Phi}_{\mathbf{k}}|^{2}
    -\frac{1}{2}a^{3}\left(M^{2}+\frac{k^{2}}{a^{2}}\right)|\Phi_{\mathbf{k}}|^{2}\right]~.
    \end{align}
    It is clear from \ref{actionforscalarinfrw} that the quantum scalar field in an expanding background can be reduced  to a bunch of harmonic oscillators with \textit{time-dependent mass} $m_{\mathbf{k}}(t)$ and frequency $\omega_{\mathbf{k}}(t)$ with
    \begin{align}\label{mass_freq_frw}
    m_{\mathbf{k}}(t)=a^3(t); \qquad \omega_{\mathbf{k}}^2(t)=M^2+\frac{k^2}{a^2}~.
    \end{align}
While such an approach is extensively used in the literature, \textit{it is obviously non-covariant} (i.e., it is based on a special choice of coordinates on the manifold) and, of course, all the results obtained by such a method will inherit this non-covariance.    
    In fact, even time dependence of the mass and frequency, is far from being unique for this system. If, instead of the geodesic time $t$, we decide to use the conformal time $\eta$ given by $d\eta=dt/a(t)$, the Lagrangian will become
    \begin{align}\label{actioneta}
    L= \left[\frac{a^2}{2}|\Phi^{'}_{\bf k}|^{2}
    -\frac{a^2}{2}\left(k^{2}+a^{2}M^{2}\right)|\Phi _{\bf k}|^{2}\right]~,
    \end{align}
    where `prime' denotes derivative with respect to $\eta$. 
    (In arriving at this Lagrangian, we have omitted a total time derivative which is irrelevant for the dynamics). 
    Hence, we again end up with a set of independent harmonic oscillators, one for each $\mathbf{k}$ mode, but with time-dependent mass and frequency given by
    \begin{align}\label{mass_and_freq_eta}
    m_{\mathbf{k}}(\eta)=a^2(\eta); \qquad  \omega_{\mathbf{k}}^2(\eta)=k^2+a^2M^2~.
    \end{align} 
    So clearly, the explicit form of time dependence of the mass and frequency of the harmonic oscillator depends on the choice of the time coordinate.
    
    In \ref{mass_and_freq_eta} and \ref{mass_freq_frw} we have the description in terms of normal oscillators with positive mass and real frequency. It is, however, straightforward to convert them into an inverted oscillator.   
    In this case  the above transformation corresponds to a change of variable from $\Phi_{\mathbf{k}}$ to $\Theta_{\mathbf{k}}=a^{3/2}\Phi_{\mathbf{k}}$ and the time-dependent frequency-squared $\Omega_{\mathbf{k}}^2(t)$ of the new oscillator $\Theta_{\mathbf{k}}$ becomes 
    \begin{align*}
    \Omega_{\mathbf{k}}^2(t)=M^2+\frac{k^2}{a^2}-\frac{3 \ddot{a}}{2 a}-\frac{3 \dot{a}^2}{4 a^2}
    \end{align*}
    In particular, the new frequency $\Omega^2_k$ when the scale factor $a$ corresponds to de Sitter and a power-law expansion takes the following form.
    \begin{align}\label{new_freq_universes}
    \Omega^2_{\mathbf{k}}(t)=
    \begin{cases}
    M^2+\frac{k^2}{a^2}-\frac{9 H^2}{4},&\textrm{when $a=e^{Ht}$ and}\\
    M^2+\frac{k^2}{a^2}+\frac{9H_0^2}{4na^{2/n}}(\frac{2}{3}-n)&\textrm{when $a=(t/t_0)^n$}
    \end{cases}
    \end{align}
    where, $H_0=\dot{a}(t_0)/a(t_0)$. When $a\propto t^n$, $\Omega_{\mathbf{k}}^2$ always remains positive for $n\leq2/3$ but can become negative when $n> 2/3$. On the other hand, in the case of de Sitter, $\Omega^2_{\mathbf{k}}$ always remains positive for scalar fields with mass $M\geq3H/4$. When $M<3H/4$, the sign of $\Omega_{\mathbf{k}}^2$ changes when the scale factor equals a critical value given by $a_c=2k/\sqrt{9H^2-4M^2}$. 
    
    In the context of a scalar field in FRW background, for the choice $f=1/a$, the action given in \ref{actionforscalarinfrw} in terms of the new variable $\vartheta_{\mathbf{k}}=a\Phi_{\mathbf{k}}$ and the new time coordinate $\eta$ defined by $ad\eta=dt$ reduces to 
    \begin{align}\label{mukhanov_sasaki}
    \mathcal{A}=\int d\eta \int \frac{d^{3}\mathbf{k}}{(2\pi)^{3}}\left[\frac{1}{2}|\vartheta^{'}_{\bf k}|^{2}
    -\frac{1}{2}\left(k^{2}+a^{2}M^{2}-\frac{a''}{a}\right)|\vartheta _{\bf k}|^{2} \right]
    -\frac{1}{2}\int d\eta \int \frac{d^{3}\mathbf{k}}{(2\pi)^{3}}\frac{d}{d\eta}\left(\frac{a'}{a}|\vartheta _{\bf k}|^{2}\right)
    \end{align} 
    Ignoring the total time derivative, the Lagrangian for the mode labelled by $\mathbf{k}$ is given by:
    \begin{align}\label{mukhanov_sasaki1}
    L=\left[\frac{1}{2}|\vartheta^{'}_{\bf k}|^{2}
    -\frac{1}{2}\left(k^{2}+a^{2}M^{2}-\frac{a''}{a}\right)|\vartheta _{\bf k}|^{2} \right]
    \end{align} 
    The new harmonic oscillator mode $\vartheta_{\mathbf{k}}$ is just the Mukhanov-Sasaki variable \cite{mukhanov1992theory,mukhanov,sasaki1986large,sasaki} familiar in the context of inflationary physics  and the time coordinate $\eta$ is just  the conformal time. 
    In the case of a massless scalar field, the relevant squared frequency is given by $\tilde{\Omega}^2_\mathbf{k}=k^2-(a''/a)$. The point of transition when this frequency changes from real to imaginary is around the epoch at which a particular mode crosses the Hubble radius. When the modes are smaller than the Hubble radius they are represented by an oscillator with real frequency while the modes are outside the Hubble radius they are represented by an inverted oscillator, that is an oscillator with imaginary frequency. So this familiar situation in inflationary cosmology is just a special case of the switching of the harmonic oscillator representation described by our transformation with $Q_f =q/f(t)$ and $dt = (mf^2) d\eta$. This links our transformation to a more familiar context discussed in the literature.

    In the context of the general oscillator system $Q_f$ we will again find that the frequency-squared $\tilde{\Omega}^2$ may change sign even when $\omega^2(t)$ is always positive. The sign of $\tilde{\Omega}^2$ is positive/negative whenever $\{\ddot{f}+(\dot{m}/m)\dot{f}+\omega^2f\}$ is positive/negative. For the harmonic oscillator modes $\vartheta_{\mathbf{k}}$ in \ref{mukhanov_sasaki}, the new frequencies $\tilde{\Omega}^2_\mathbf{k}$ for the de Sitter and power-law FRW universe are given by
    \begin{align}\label{newOmega}
    \tilde{\Omega}_k^2=
    \begin{cases}
    k^2+(M^2-2H^2)a^2,&\textrm{when $a=e^{Ht}$ and}\\
    k^2+a^2M^2+\frac{(1-2n)H_0^2}{n}a^{\frac{2(n-1)}{n}},&\textrm{when $a=\left(t/t_0\right)^{n}$}
    \end{cases}
    \end{align}
    For a power-law scale factor, we see that the sign of $\tilde{\Omega}_{\mathbf{k}}^2$ is always positive when $n\leq1/2$, \textit{which is different from the range for $n$ which we found from} \ref{new_freq_universes}.  For the de Sitter case, $\tilde{\Omega}_{\mathbf{k}}^2$ is always positive if the mass of the scalar field $M\geq\sqrt{2}H$ which is also different from the range we found in \ref{new_freq_universes}. When $M<\sqrt{2}H$, the sign of $\tilde{\Omega}_\mathbf{k}^2$ for the de Sitter changes when the scale factor equals $\tilde{a}_c=k/\sqrt{M^2-2H^2}$. So this effect depends not only on the form of the dynamical variable used but also on the time coordinate we are working with.
    
    Up to this point we have only discussed a \textit{mathematical} correspondence between (i) an oscillator system, discussed in the main body of the paper and (ii) a mode of the scalar field labelled by $\mathbf{k}$. One can proceed further purely formally, treating each degree of freedom labelled by $\mathbf{k}$, as harmonic oscillator. In the familiar case of a massless scalar field in a de Sitter background described in terms of  the conformal time $\eta$,
	 the Mukhanov-Sasaki variables $\vartheta_{\bf k}$, satisfies a time-dependent harmonic oscillator equation with unit mass and frequency given by \ref{newOmega} with $M=0$.
	The solution to this equation  with the initial condition $\mu_{\mathbf{k}}\propto e^{ik\eta}$ as $\eta\rightarrow-\infty$ is given by
	\begin{align}
	\mu_{\mathbf{k}}(\eta)=\frac{e^{i \eta  k} (-1+i \eta k)}{\sqrt{2} \eta  k^{3/2}}
	\end{align}
	where, the Wronskian condition $i(\mu_{\mathbf{k}}(\mu_{\mathbf{k}}^{*})'-\mu_{\mathbf{k}}'\mu_{\mathbf{k}}^*)=1$ is also imposed. 
	One can then compute, for example,
	the
	 expectation value of the operator $H_{\mathbf{k}}=P_{\vartheta_{\mathbf{k}}}^2/2+(k^2-2/\eta^2)\vartheta_{\mathbf{k}}^2$, where $P_{\vartheta_{\mathbf{k}}}$ is the momentum operator corresponding to $\vartheta_{\mathbf{k}}$. We will get:
	\begin{align}
	\braket{H_{\mathbf{k}}}=\frac{1}{2}\left(|\mu_{\mathbf{k}}|^2+\Omega_{\mathbf{k}}^2|\mu_{\mathbf{k}}|^2\right)=\frac{k}{2}\left(1-\frac{1}{2 \eta ^4 k^4}-\frac{1}{ \eta
		^2 k^2}\right)
		\label{aaa}
	\end{align}
	We can see that, as expected, $\braket{H_{\mathbf{k}}}\rightarrow k/2$ as $\eta\rightarrow-\infty$ showing that the oscillator was at the ground state in the asymptotic past. The sign of $\Omega_{\mathbf{k}}^2=(k^2-2/\eta^2)$ changes from positive to negative after a critical time $\eta_c\equiv-\sqrt{2}/k$. Obviously the standard approach to defining the ``particle number'' will not work after this critical time. But expectation value of $H_{\mathbf{k}}$, however, behaves smoothly in the neighbourhood of $\eta_c$, as can be seen from a simple Taylor series expansion.
	As time evolves further there is another  critical moment $\eta_H=-k^{-1}\sqrt{(1+\sqrt{3})/2}$ at which the sign of $\braket{H_{\mathbf{k}}}$ itself switches to negative values. If we compute the expectation value of the ``number'' operator using from \ref{newnumber},  we will find that,
	 as we approach $\eta_c$ from $\eta<\eta_c$, this expectation value is real and  \textit{diverges} as $1/\Omega_{\mathbf{k}}$. On the other hand, as we approach $\eta_c$ from $\eta>\eta_c$, we can see that $n_{\mathbf{k}}$ is a complex number with a constant real part, namely $-1/2$ and an imaginary part that diverges as  $1/\Omega_{\mathbf{k}}$. These are special case of the general discussion we provided in the body of the paper.
	
	In contrast to the Mukhanov-Sasaki variables, the  behaviour  associated with the Fourier modes $\Phi_{\bf k}$ of the massless scalar field presents no surprises.
	These  Fourier modes $\Phi_{\bf k}$ of the massless scalar field  obeys a time-dependent harmonic oscillator equation with a squared-frequency $\omega^2_{\mathbf{k}}$ which is \textit{always positive}. Further, $\xi_{\mathbf{k}}=\mu_{\mathbf{k}}/a$ is a solution to the equation of motion for $\Phi_{\bf k}$ with the initial condition that the mode functions $\xi_{\mathbf{k}}$ are positive frequency solutions near $\eta\rightarrow-\infty$ (defining the usual Bunch-Davies vacuum). The expectation value of the Hamiltonian $H_{\mathbf{k}(2)}=P_{\Phi_{\mathbf{k}}}^2/(2m_{\mathbf{k}})+\omega_{\mathbf{k}}^2\Phi_{\mathbf{k}}^2$, where $P_{\Phi_{\mathbf{k}}}$ is the momentum operator associated with $\Phi_{\mathbf{k}}$, turns out to be 
	\begin{align}
	\braket{H_{\mathbf{k}(2)}}=\frac{k}{2}\left(1+\frac{1}{2k^2\eta^2}\right)
	\end{align}
	which, in contrast to $\braket{H_{\mathbf{k}}}$ in \ref{aaa}, is strictly positive. The ``particle number'' $N_{\mathbf{k}}$ corresponding to the $\Phi_{\mathbf{k}}$ can be found to be
	\begin{align}
	N_{\mathbf{k}}=\frac{1}{4k^2\eta^2}
	\end{align}
	which, in contrast to $n_{\mathbf{k}}$, is always real and is monotonically increasing in a smooth manner. In particular, at the moment $\eta_c$ when the particle number associated with the inverted oscillator representation diverges, the particle number associated with the normal oscillator is $N_{\mathbf{k}}(\eta_{c})=1/8$, which is a finite value. 
	
	Thus, as long as the back-reaction of the scalar field on geometry is ignored, the \textit{mathematical} description given above closely parallels the discussion in the first part of the paper. This is because different oscillator modes have  decoupled from each other and we can concentrate on any given oscillator mode and study its dynamics. 
	
The \textit{physical interpretation} of the system, however, raises new issues because the quantum field requires infinite number of oscillators for its description. As a cautionary note to the reader, we list these issues, which are well-known in the literature.
	(a) The expectation value $\langle H_{\mathbf{k}} \rangle$ cannot be used to compute the total energy of the quantum filed  because the sum over all modes  diverges. In free field theories described in inertial coordinates, one can resort to a normal-ordering prescription to make the renormalized energy of the vacuum is exactly zero. This  prescription, however, fails in non-trivial backgrounds, because normal ordering is ambiguous, not covariant and not always compatible
with causality. So the naive interpretation of individual modes fails.
	(b) Similar comments apply to the definition of particle number in curved backgrounds. To begin with, we have no unambiguous way of defining the positive frequency modes, vacuum state and thus ``particles'' in a time dependent background. Also in a general, curved, background it is not always possible to find a  vacuum
state which is an instantaneous eigenstate of the Hamiltonian and ultraviolet-regular (see e.g. \cite{Fulling1979}). So it is not easy to define ``instantaneous particle number'' in general.
(c) Given these ambiguities, it is also difficult to generalize the back-reaction prescription for a single mode to the full quantum field because, again, naive summation over the modes will lead to a divergence. The only context in which such a prescription will work is when we have a preferred choice of oscillators and a preferred vacuum state. But such a choice will not be generally covariant and it is not clear whether one can trust the results of a back-reaction calculation, which is not based on a generally covariant prescription.

We thought it is appropriate to add these caveats explicitly because one often resorts to non-covariant methodology in the study of quantum fields in expanding backgrounds.

	\section{Relation to over-damping of an oscillator}\label{rel_to_overdamp}
	
	The switching of sign of $\Omega^2$ is related to a more familiar phenomena known in the context of oscillators.
	The switching of the sign of $\Omega^2$ can be interpreted as a transition of the $q$ system from an \textit{under-damped} to \textit{over-damped} oscillator. To see this, let us recall the equation of motion for a damped harmonic oscillator.
	\begin{align}\label{damped}
	\ddot{x}+2\gamma\dot{x}+\omega_0^2x=0
	\end{align}   
	where, $\gamma$ and $\omega_0^2$ are positive constants. The general solution for \ref{damped} is given by
	\begin{align}\label{damped_solution}
	x(t)=e^{-\gamma t}\left(C_+e^{i\Omega_0 t}+C_-e^{-i\Omega_0 t}\right)	
	\end{align}
	where, 
	\begin{align}\label{Omega_0}
	\Omega_0^2=\omega_0^2-\gamma^2
	\end{align}
	and, $C_-$ and $C_+$ are constants. Therefore, \ref{damped} has oscillating solutions when $\Omega_0^2>0$, in which case we say that the oscillator is \textit{under-damped}. On the other hand, when $\Omega_0^2<0$, \ref{damped} has only decaying solutions and we say that the oscillator is \textit{over-damped}. Let us now transform to a new variable $X=e^{\gamma t}x$, so that \ref{damped} becomes
	\begin{align}\label{damped_X}
	\ddot{X}+\Omega_0^2X=0.
	\end{align}
	\ref{damped_X} describes a simple harmonic oscillator when $\Omega_0^2>0$ and an inverted harmonic oscillator when $\Omega_0^2<0$. Therefore, the damped oscillator system $x$ is \textit{under-damped}/\textit{over-damped} depending on whether $X$ describes a simple/inverted harmonic oscillator. Let us come back to the system given by \ref{actiontimedependent}. The equation of motion for $q$ is given by
	\begin{align}\label{eom_for_q_2}
	\ddot{q}+\frac{\dot{m}}{m}\dot{q}+\omega^2q=0
	\end{align}  
	When $\dot{m}/m>0$, \ref{eom_for_q} can be interpreted as the time-dependent generalization of \ref{damped}, with a time-dependent damping factor given by $\gamma(t)=\dot{m}/(2m)$. The transformation from $x\rightarrow X$ generalizes to
	\begin{align}\label{Q_to_q_sqrt_m}
	Q=\exp\left(\int\gamma(t')dt'\right)q\propto \sqrt{m}q
	\end{align}
	which is similar to the transformation introduced in \ref{action_negative_freq} and the time-dependent generalization of \ref{Omega_0} is precisely given by \ref{mass_sqrt_frequency}. Hence, the transition of system $Q$ from a simple to inverted oscillator can be interpreted as a transition of $q$ oscillator from under-damped to over-damped phase.
	
	\section{Solving the Schr\"{o}dinger equation for a quantum mechanical time-dependent oscillator}\label{solving_se1}
	
	The Schr\"{o}dinger equation for a time-dependent oscillator is given by
	\begin{equation}
	\label{se2}
	i\partial_t\psi(t,q)=-\frac{1}{2m(t)}\partial_{q}^2\psi(t,q)+\frac{m(t)\omega(t)^2}{2}\psi(t,q)
	\end{equation} 
	This equation can be solved, for our purpose, by the following Gaussian ansatz (see, e.g., \cite{mahajan2008particle})
	\begin{align}
	\psi(q,t)=\mathcal{N}(t)\exp\left(-R(t)q^2\right).
	\end{align}
	The function $\psi$ satisfies the Schr\"{o}dinger equation \ref{se2} if
	\begin{align}
	i\frac{\dot{\mathcal{N}}}{\mathcal{N}}&=\frac{R}{m}\\\label{Reqn}
	i\dot{R}&=\frac{2R^2}{m}-\frac{1}{2}m\omega^2
	\end{align}
	It is convenient at this point to define $R=-(im/2)(\dot{\xi}/\xi)$ so that \ref{Reqn} becomes
	\begin{align}\label{eom_for_xi}
	\ddot{\xi}+\frac{\dot{m}}{m}\dot{\xi}+\Omega^2\xi=0
	\end{align}
	If the Gaussian ansatz $\psi$ corresponds to the `in-vacuum' state $\ket{\textrm{in}}$ defined at $t=t_1$ then, $\xi$ should satisfy the initial condition $\dot{\xi}(t_1)=i\omega(t_1)$.  Therefore, the normalized wave function $\braket{q|\textrm{in}}\equiv\psi_{in}$ is given by
	\begin{align}
	\psi_{in}(q,t)=\exp\left(-\int_{t_1}^{t}\frac{\dot{\xi}}{2\xi}dt+\frac{im\dot{\xi}}{2\xi}q^2\right)
	\end{align}
	with the initial condition that $\xi\approx e^{i\omega t}$ as $t\rightarrow t_1$.
	
	\section{Transformation of kernel and wave function under $q\rightarrow Q_f$}\label{kernel_and_wavefn}
	Let us take the following ansatz by which the wave function for the $q$ system $\psi$ and that of the $Q_f$ system $\Psi$ are related
	\begin{align}
	\psi=A\exp(Bq^{2})\Psi
	\end{align}
	with $A$ and $B$ depending on time. Thus, from \ref{se1}, we obtain,
	\begin{align}
	i\frac{\partial \psi}{\partial t}&=i\frac{\partial A}{\partial t}\exp(Bq^{2})\Psi +iA\exp(Bq^{2})\frac{\partial \Psi}{\partial t}
	-iA\exp(Bq^{2})\frac{\partial \Psi}{\partial Q_f}\left(\frac{q\dot{f}}{f^2}\right)
	+iA\exp(Bq^{2})\Psi q^{2}\frac{\partial B}{\partial t}
	\\
	\frac{\partial \psi}{\partial q}&=A\exp(Bq^{2})\frac{\partial \Psi}{\partial Q_f}f^{-1}+2AB\exp(Bq^{2})q\Psi
	\\
	\frac{\partial ^{2}\psi}{\partial q^{2}}&=Af^{-2}\exp(Bq^{2})\frac{\partial ^{2}\Psi}{\partial Q_f^{2}}
	+4ABQ_f\exp(Bq^{2})\frac{\partial \Psi}{\partial Q_f}+4AB^{2}\left(Q_f\right)^{2}\exp(Bq^{2})\Psi
	+2AB \exp(Bq^{2})\Psi
	\end{align}
	Hence the Schr\"{o}dinger equation for $\psi$ yields,
	\begin{align}
	i\frac{\partial \ln A}{\partial t}\Psi&+\frac{i}{mf^2}\frac{\partial \Psi}{\partial \eta}-i \frac{\partial \Psi}{\partial Q_f} \frac{Q_ff'}{mf^3}+i\left(Q_ff\right)^{2}\Psi \frac{\partial B}{\partial t}
	\nonumber
	\\
	&=-\frac{1}{2m}\Big[f^{-2}\frac{\partial ^{2}\Psi}{\partial Q_f^{2}}+4BQ_f\frac{\partial \Psi}{\partial Q_f}+4B^{2}
	\left(Q_f\right)^{2}+2B \Big]+m\omega ^{2}Q_f^{2}f^2
	\end{align}
	Using the differential equation for $\Psi$ and the expression of $\tilde{\Omega} ^{2}$ in terms of other parameters in the model we obtain,
	\begin{align}
	B=i\frac{f'}{2f^3}=\frac{im\dot{f}}{2f}
	\end{align}
	by equating coefficient of $\partial \Psi/\partial Q_f$. Further equating the terms with out any powers of $Q_f$ on both sides we obtain,
	\begin{align}
	i\frac{\partial \ln A}{\partial t}=-\frac{B}{m};\qquad A=1/\sqrt{f}
	\end{align}
	Hence we obtain the relation between the two wave functions given in \ref{Eq_psi_phi}. 
	
	To obtain the relation between the two propagation kernels,  recall that the propagation equation for the wave function is given by
	\begin{align}\label{define_kernel}
	\psi(t_2,q_2)=\int dq_{1}~K_{q}(t_2,q_2;t_{1},q_{1})\psi(t_{1},q_{1})
	\end{align}
	Using \ref{Eq_psi_phi} and the definition of $K_Q$, we can rewrite this as
	\begin{align}\label{kernel_substituted}
	\psi(t_2,q_2)=\left[\frac{1}{\sqrt{f}}\exp\left(i\frac{m\dot{f}}{2f}q_2^{2}\right)\right]\bigg\rvert_{t=t_2}\int dQ_{f1}~K_{Q}(\eta_2,Q_{f2};\eta_{1},Q_{f1})\Psi(\eta_{1},Q_{f1})
	\end{align}
	On the other hand, using \ref{Eq_psi_phi}, we can replace $\psi$ in \ref{define_kernel} with $\Phi$ to get 
	\begin{align}
	\int dq_{1}~\left[\frac{1}{\sqrt{f}}\exp\left(i\frac{m\dot{f}}{2f}q_1^{2}\right)\right]\bigg\rvert_{t=t_1} &K_{q}(t_2,q_2;t_{1},q_{1})\Psi(\eta_{1},Q_{f1})
	\nonumber
	\\
	&=\left[\frac{1}{\sqrt{f}}\exp\left(i\frac{m\dot{f}}{2f}q_2^{2}\right)\right]\bigg\rvert_{t=t_2}\int dQ_{f1}~K_{Q}(\eta_2,Q_{f2},\eta_{1},Q_{f1})\Phi(\eta_{1},Q_{f1})
	\end{align}
	Since, $f_{1}dQ_{f1}=dq_{1}$, we see that the Kernels must be connected by \ref{K_q_to_Q}.
	
	\section{Derivation of \ref{eom_for_z}}\label{derivationforz}
	From \ref{bogolforinver1} we can see that the Bogoliubov coefficients satisfy the following differential equation.
	\begin{align}\label{derivativeA}
	\dot{\mathcal{A}}&=\frac{1}{2}\left(\frac{\dot{\omega}}{\omega}+\frac{\dot{m}}{m}\right)\mathcal{B}e^{2i\rho}\\\label{derivativeB}
	\dot{\mathcal{B}}&=\frac{1}{2}\left(\frac{\dot{\omega}}{\omega}+\frac{\dot{m}}{m}\right)\mathcal{A}e^{-2i\rho}
	\end{align}
	where, $\dot{\rho}=\omega$. It follows from the definition that the time derivative of $z$ is given by
	\begin{align}\label{zdot}
	\dot{z}=-\frac{\mathcal{B}^* 
		\dot{\mathcal{A}}^*}{(\mathcal{A}^*)^2}e^{-2 i \rho}+\frac{
		\dot{\mathcal{B}^*}}{\mathcal{A}^*}e^{-2 i \rho}-\frac{2 i \mathcal{B}^* 
		\omega}{\mathcal{A}^*}e^{-2 i \rho}
	\end{align} 
	Using, \ref{derivativeA}, \ref{derivativeB} and the definition of $z$ in \ref{zdot}, we get the following differential equation.
	\begin{align}
	\dot{z}+2i\omega z+\frac{1}{2}\left(\frac{\dot{\omega}}{\omega}+\frac{\dot{m}}{m}\right)(z^2-1)=0
	\end{align}
	
	\section{A brief comparison of `in-out' and `in-in' approaches}\label{compare_ii_oi}
	We have \ref{BkReac_overview_intuitive} we discussed the `in-in' approach to quantum back-reaction for a $C-q$ system with quadratic coupling. However, there is another popular prescription for back-reaction called the `in-out' approach. In this approach, \ref{bk_reac} will be replaced by
	\begin{align}\label{bk_reac_io}
	\mathcal{M}\ddot{C}+\mathcal{M}\partial_C V(C)-\frac{\partial_Cm(C)}{2m^2(C)}\frac{\braket{\textrm{out}|\hat{p}^2|\textrm{in}}}{\braket{\textrm{out}|\textrm{in}}}+\frac{1}{2}\partial_C\left(m(C)\omega^2(C)\right)\frac{\braket{\textrm{out}|\hat{q}^2|\textrm{in}}}{\braket{\textrm{out}|\textrm{in}}}=0
	\end{align}
	where, the `out-vacuum' $\ket{\textrm{out}}$ is defined by $a_-(T)\ket{\textrm{out}}=0$ where $t=T$ is an appropriate choice of the final time.
	The conservation of energy equation \ref{conserve_quad}  will be replaced by
	\begin{align}
	\label{quadratice_energy_inout}
	\textrm{In-out approach:\,\,\,}\frac{d}{dt}\left[\frac{M}{2}\dot{C}^{2}+V(C)+\left(n+\frac{1}{2}-\underbrace{\frac{\mathcal{B}(T)}{\mathcal{A}^*(T)}\mathcal{A}^*(t)\mathcal{B}^*(t)}_{\equiv \varepsilon_{ext}/\omega}\right)\omega\right]=0
	\end{align}
	where $\mathcal{A}$ and $\mathcal{B}$ are as defined in \ref{A_and_B}. For contrast, from \ref{quadratice_energy_inout} we see that, in the `in-out' formalism, due to the presence of the extra term $\varepsilon_{ext}$, the corresponding drain $\Delta E_{c}^{in-out}$ of the classical system $C$ is given by 
	\begin{align}
	\Delta E_{c}^{in-out}=-\frac{1}{2}\left(\omega(T)-\omega(t_1)\right)
	\end{align}
	The production of an average of $\mathcal{B}\mathcal{\tilde{B}}$ number of `$q$-particle' seems to have not contributed a proportional amount of energy to the classical system in the in-out approach. This is clearly an undesirable feature since, in general, the produced particles do carry a non-zero energy which should be accounted for in a consistent theory of back-reaction. Moreover, the back-reaction equation corresponding to \ref{quadratice_energy_inout} is non-local in time and has non-causal behaviour. On the other hand, the `in-in' back-reaction approach given in \ref{conserve_quad} is clearly devoid of these shortcomings.
	
	\section{Classical dynamics and Energy function for $(C,q)$ and $(C,Q)$ systems}\label{classical_dyn}
	In this section we briefly review the classical dynamics of the coupled systems $(C,q)$ and  $(C,Q)$, with special emphasis on the conservation of energy equations. We start with the Lagrangian $L(C,\dot{C},q,\dot{q})$ for the $(C,q)$ given by,
	\begin{align}
	L_1(C,\dot{C},q,\dot{q})=\frac{1}{2}\mathcal{M}\dot{C}^{2}-\mathcal{M}V(C)+\frac{1}{2}m(C)\dot{q}^{2}-\frac{1}{2}m(C)\omega ^{2}(C)q^{2}
	\end{align}
	Then we obtain,
	\begin{align}
	\frac{dL_1}{dt}&=\frac{\partial L_1}{\partial C}\dot{C}+\frac{\partial L_1}{\partial \dot{C}}\ddot{C}+\frac{\partial L_1}{\partial q}\dot{q}+\frac{\partial L_1}{\partial \dot{q}}\ddot{q}
	\nonumber
	\\
	&=\frac{d}{dt}\left(\frac{\partial L_1}{\partial \dot{C}}\dot{C}+\frac{\partial L_1}{\partial \dot{q}}\dot{q}\right)
	+\left\{\frac{\partial L_1}{\partial C}-\frac{d}{dt}\left(\frac{\partial L_1}{\partial \dot{C}}\right)\right\}\dot{C}
	+\left\{\frac{\partial L_1}{\partial q}-\frac{d}{dt}\left(\frac{\partial L_1}{\partial \dot{q}}\right)\right\}\dot{q}
	\end{align}
	Therefore, when the Euler-Lagrange equations for both $C$ and $q$ are satisfied, one obtains the following identity
	\begin{align}
	\frac{d}{dt}\left(\frac{\partial L_1}{\partial \dot{C}}\dot{C}+\frac{\partial L_1}{\partial \dot{q}}\dot{q}-L_1\right)=0
	\end{align}
	thereby defining the associated energy function. The associated energy function would correspond to,
	\begin{align}
	E_{C,q}&=\dot{C}\frac{\partial L_1}{\partial \dot{C}}+\dot{q}\frac{\partial L_1}{\partial \dot{q}}-L_1
	\nonumber
	\\
	&=\mathcal{M}\dot{C}^{2}+m(C)\dot{q}^{2}-\left(\frac{1}{2}\mathcal{M}\dot{C}^{2}-\mathcal{M}V(C)+\frac{1}{2}m(C)\dot{q}^{2}-\frac{1}{2}m(C)\omega ^{2}(C)q^{2}\right)
	\nonumber
	\\
	&=\frac{1}{2}\mathcal{M}\dot{C}^{2}+\mathcal{M}V(C)+\frac{1}{2}m(C)\dot{q}^{2}+\frac{1}{2}m(C)\omega ^{2}(C)q^{2}
	\end{align}
	Further, by our previous argument setting $dE_{C,q}/dt=0$ will provide the Euler-Lagrange equations. To check that this is indeed the case, note that the Euler-Lagrange equation for $q$ correspond to 
	\begin{align}
	m\ddot{q}+\dot{m}\dot{q}+m\omega ^{2}q&=0
	\\
	\mathcal{M}\ddot{C}+\mathcal{M}\frac{\partial V}{\partial C}-\frac{1}{2}\frac{\partial m}{\partial C}\dot{q}^{2}+\frac{1}{2}\frac{\partial (m\omega ^{2})}{\partial C}q^{2}&=0
	\end{align}
	While,
	\begin{align}
	\frac{dE_{C,q}}{dt}&=\mathcal{M}\dot{C}\ddot{C}+\mathcal{M}\frac{\partial V}{\partial C}\dot{C}
	+\frac{1}{2}\frac{\partial m}{\partial C}\dot{C}\dot{q}^{2}+\frac{1}{2}\frac{\partial (m\omega ^{2})}{\partial C}\dot{C}q^{2}
	+m\dot{q}\ddot{q}+m\omega ^{2}q\dot{q}
	\nonumber
	\\
	&=\mathcal{M}\dot{C}\ddot{C}+\mathcal{M}\frac{\partial V}{\partial C}\dot{C}
	+\frac{1}{2}\frac{\partial m}{\partial C}\dot{C}\dot{q}^{2}+\frac{1}{2}\frac{\partial (m\omega ^{2})}{\partial C}\dot{C}q^{2}
	+m\dot{q}\ddot{q}+m\omega ^{2}q\dot{q}
	\nonumber
	\\
	&=\mathcal{M}\dot{C}\ddot{C}+\mathcal{M}\frac{\partial V}{\partial C}\dot{C}
	+\frac{1}{2}\frac{\partial m}{\partial C}\dot{C}\dot{q}^{2}+\frac{1}{2}\frac{\partial (m\omega ^{2})}{\partial C}\dot{C}q^{2}
	+\dot{q}\{-\dot{m}\dot{q}-m\omega ^{2}q\}+m\omega ^{2}q\dot{q}
	\end{align}
	which will coincide with the equation of motion for $C$ as expected. So far, all this is elementary classical mechanics. 
	
	Let us transform the above Lagrangian to another one with a new variable $Q$, such that $Q=\sqrt{m}~q$, thus the Lagrangian would become,
	\begin{align}
	L_1&=\frac{1}{2}\mathcal{M}\dot{C}^2-\mathcal{M}V(C)+\frac{1}{2}m\dot{q}^{2}-\frac{1}{2}m\omega ^{2}q^{2}
	\nonumber
	\\
	&=\frac{1}{2}\mathcal{M}\dot{C}^2-\mathcal{M}V(C)+\frac{1}{2}m\left(\frac{\dot{Q}}{\sqrt{m}}-\frac{1}{2}\frac{\dot{m}Q}{m\sqrt{m}}\right)^{2}-\frac{1}{2}\omega ^{2}Q^{2}
	\nonumber
	\\
	&=\frac{1}{2}\mathcal{M}\dot{C}^2-\mathcal{M}V(C)+\frac{1}{2}\dot{Q}^{2}-\frac{\dot{m}}{2m}Q\dot{Q}+\frac{\dot{m}^{2}}{8m^{2}}Q^{2}-\frac{1}{2}\omega ^{2}Q^{2}
	\nonumber
	\\
	&=\frac{1}{2}\mathcal{M}\dot{C}^2-\mathcal{M}V(C)+\frac{1}{2}\dot{Q}^{2}-\frac{1}{2}\left(\omega ^{2}+\frac{\dot{m}^{2}}{4m^{2}}-\frac{\ddot{m}}{2m}\right)Q^{2}-\frac{d}{dt}\left(\frac{\dot{m}}{4m}Q^{2}\right)
	\end{align}
	Thus ignoring the total derivative and keeping in mind that both $m$ and $\omega$ are functions of time through $C(t)$, it follows that, the total Lagrangian becomes,
	\begin{align}
	L_2(C,\dot{C},\ddot{C},Q,\dot{Q})=\frac{1}{2}\mathcal{M}\dot{C}^{2}-\mathcal{M}V(C)
	+\frac{1}{2}\dot{Q}^{2}-\frac{1}{2}\left(\omega ^{2}+\frac{m'^{2}}{4m^{2}}\dot{C}^{2}-\frac{m''}{2m}\dot{C}^{2}-\frac{m'}{2m}\ddot{C}\right)Q^{2}
	\end{align}
	For such a Lagrangian, we obtain, 
	\begin{align}
	\frac{dL_2}{dt}&=\frac{\partial L_2}{\partial C}\dot{C}+\frac{\partial L_2}{\partial \dot{C}}\ddot{C}+\frac{\partial L_2}{\partial \ddot{C}}\dddot{C}+\frac{\partial L_2}{\partial Q}\dot{Q}+\frac{\partial L_2}{\partial \dot{Q}}\ddot{Q}+
	\nonumber
	\\
	&=\frac{d}{dt}\left\{\frac{\partial L_2}{\partial \dot{C}}\dot{C}+\frac{\partial L_2}{\partial \ddot{C}}\ddot{C}
	-\frac{d}{dt}\left(\frac{\partial L_2}{\partial \ddot{C}}\right)\dot{C}+\frac{\partial L_2}{\partial \dot{Q}}\dot{Q}\right\}
	\nonumber
	\\
	&+\left\{\frac{\partial L_2}{\partial C}-\frac{d}{dt}\left(\frac{\partial L_2}{\partial \dot{C}}\right) 
	+\frac{d^{2}}{dt^{2}}\left(\frac{\partial L_2}{\partial \ddot{C}}\right)\right\}\dot{C}
	+\left\{\frac{\partial L_2}{\partial Q}-\frac{d}{dt}\left(\frac{\partial L_2}{\partial \dot{Q}}\right)\right\}\dot{Q}
	\end{align}
	Therefore the Euler-Lagrange equations for $Q$ reads,
	\begin{align}
	\ddot{Q}+\left(\omega ^{2}+\frac{\dot{m}^{2}}{4m^{2}}-\frac{\ddot{m}}{2m}\right)Q&=0
	\end{align}
	The Energy function becomes,
	\begin{align}
	E_{C,Q}&=\frac{\partial L_2}{\partial \dot{C}}\dot{C}+\frac{\partial L_2}{\partial \ddot{C}}\ddot{C}
	-\frac{d}{dt}\left(\frac{\partial L_2}{\partial \ddot{C}}\right)\dot{C}+\frac{\partial L_2}{\partial \dot{Q}}\dot{Q}-L_2
	\nonumber
	\\
	&=\dot{C}\left[\mathcal{M}\dot{C}-\frac{m'^{2}}{4m^{2}}\dot{C}+\frac{m''}{2m}\dot{C}\right]Q^{2}
	+\ddot{C}\left[\frac{m'}{4m}\right]Q^{2}-\dot{C}\frac{d}{dt}\left(\frac{m'}{4m}Q^{2} \right)
	+\dot{Q}^{2}-\frac{1}{2}\mathcal{M}\dot{C}^{2}+\mathcal{M}V(C)
	\nonumber
	\\
	&-\frac{1}{2}\dot{Q}^{2}+\frac{1}{2}\left(\omega ^{2}+\frac{m'^{2}}{4m^{2}}\dot{C}^{2}-\frac{m''}{2m}\dot{C}^{2}-\frac{m'}{2m}\ddot{C}\right)Q^{2}
	\nonumber
	\\
	&=\frac{1}{2}\mathcal{M}\dot{C}^{2}+\mathcal{M}V(C)+\frac{1}{2}\dot{Q}^{2}+\frac{1}{2}\omega ^{2}Q^{2}+\frac{\dot{m}^{2}}{8m^{2}}Q^{2}-\frac{\dot{m}}{2m}Q\dot{Q}
	\end{align}
	Substituting $Q=\sqrt{m}q$, we obtain 
	\begin{align}
	E_{C,Q}&=\frac{1}{2}\mathcal{M}\dot{C}^{2}+\mathcal{M}V(C)+\frac{1}{2}\left(\sqrt{m}\dot{q}+\frac{1}{2}\frac{\dot{m}}{\sqrt{m}}q \right)^{2}+\frac{1}{2}m\omega ^{2}q^{2}+\frac{\dot{m}^{2}}{8m}q^{2}-\frac{\dot{m}}{2\sqrt{m}}q\left(\sqrt{m}\dot{q}+\frac{1}{2}\frac{\dot{m}}{\sqrt{m}}q \right)
	\nonumber
	\\
	&=E_{C,q}
	\end{align}
	Therefore the two energy functions are numerically the same. This is closely related to the fact that the transformation from $q$ to $Q$ is a canonical transformation generated by a generating function which is independent of time thereby making the two Hamiltonians numerically equal; we will do this explicitly at the end.)
	
	On the other hand, given the Lagrangian $L_2(C,\dot{C},\ddot{C},q,\dot{q})$ one can also make the following naive choice for the energy function, by adding the energy of the oscillator to that of the $C$ degree of freedom:
	\begin{align}
	\mathcal{E}_{C,Q}=\frac{1}{2}\mathcal{M}\dot{C}^{2}+\mathcal{M}V(C)
	+\frac{1}{2}\dot{Q}^{2}+\frac{1}{2}\left(\omega ^{2}+\frac{\dot{m}^{2}}{4m^{2}}-\frac{\ddot{m}}{2m}\right)Q^{2}
	\end{align}
	This is, of course not the conserved energy function, precisely because of the existence of $\ddot C$ in the Lagrangian.
	The difference between the above naive energy function with the actual one is
	\begin{align}
	\mathcal{E}_{C,Q}-E_{C,Q}=-\frac{\ddot{m}}{4m}Q^{2}+\frac{\dot{m}}{2m}Q\dot{Q}
	=\frac{d}{dt}\left(\frac{\dot{m}}{4m}Q^{2}\right)-\frac{\ddot{m}}{2m}Q^{2}+\frac{\dot{m}^{2}}{4m^{2}}Q^{2}
	\end{align}
	which is not constant and is not even a total time derivative (such a scenario appears in the context of gravitational action as well, see \cite{Parattu:2015gga,Chakraborty:2017zep,Chakraborty:2016yna}). For Mukhanov-Sasaki variable, for example, the difference between the two energy functions correspond to,
	a\begin{align}
	\mathcal{E}_{a,\vartheta}-E_{a,\vartheta}=-\frac{a''}{4a}\vartheta^{2}+\frac{a'}{2a}\vartheta \vartheta'
	\end{align}
	where `prime' denotes differentiation with respect to conformal time.
	
	We have not yet discussed about the equations of motion for the $C$ system, derived from the Lagrangian of the $C-Q$ system. Let us perform the above and show just like the energy functions the equations of motions are also identical (numerically) whether you use $q$ or $Q$. Variation of the higher derivative Lagrangian for the $Q$ system reads,
	\begin{align}
	0&=-\mathcal{M}\ddot{C}-\mathcal{M}\frac{\partial V}{\partial C}-\frac{1}{2}\left(\partial _{C}\omega ^{2}\right)Q^{2}
	\nonumber
	\\
	&-\left(\frac{m'm''}{4m^{2}}\right)\dot{C}^{2}Q^{2}-\left(\frac{m'^{3}\dot{C}^{2}}{4m^{3}}Q^{2} \right)
	+\left(\frac{m'^{2}}{4m^{2}}\right)\ddot{C}Q^{2}+\left(\frac{m'm''}{2m^{2}}\right)\dot{C}^{2}Q^{2}
	+\left(\frac{m'^{2}}{2m^{2}}\right)\dot{C}Q\dot{Q}
	\nonumber
	\\
	&-\left(\frac{m'''}{4m}\right)\dot{C}^{2}Q^{2}+\left(\frac{m'm''}{4m^{2}}\right)\dot{C}^{2}Q^{2}-\left(\frac{m''}{2m}\right)\ddot{C}Q^{2}
	-\left(\frac{m''}{m}\right)\dot{C}Q\dot{Q}
	\nonumber
	\\
	&+\left(\frac{m''}{2m}\right)\ddot{C}Q^{2}-\left(\frac{m'^{2}}{2m^{2}}\right)\ddot{C}Q^{2}+\left(\frac{m''}{4m}\right)\dot{C}^{2}Q^{2}
	-\left(\frac{3m'm''}{4m^{2}}\right)\dot{C}^{2}Q^{2}
	\nonumber
	\\
	&+\left(\frac{m''}{m}\right)\dot{C}Q\dot{Q}+\left(\frac{m'^{3}}{2m^{3}}\right)\dot{C}^{2}Q^{2}-\left(\frac{m'^{2}}{m^{2}}\right)\dot{C}Q\dot{Q}+\left(\frac{m'}{2m}\right)Q\ddot{Q}+\left(\frac{m'}{2m}\right)\dot{Q}^{2}
	\nonumber
	\\
	&=-\mathcal{M}\ddot{C}-\mathcal{M}\frac{\partial V}{\partial C}-\frac{1}{2}\left(\partial _{C}\omega ^{2}\right)Q^{2}
	\nonumber
	\\
	&-\left(\frac{m'^{2}}{4m^{2}}\right)\ddot{C}Q^{2}-\left(\frac{m'm''}{4m^{2}}\right)\dot{C}^{2}Q^{2}
	+\left(\frac{m'^{3}}{4m^{3}}\right)\dot{C}^{2}Q^{2}-\left(\frac{m'^{2}}{2m^{2}}\right)\dot{C}Q\dot{Q}+\frac{m'}{2m}\dot{Q}^{2}+\frac{m'}{2m}Q\ddot{Q}
	\end{align}
	Using the equation of motion for $Q$, we obtain,
	\begin{align}
	\mathcal{M}\ddot{C}+\mathcal{M}\frac{\partial V}{\partial C}-\frac{m'}{2m}\dot{Q}^{2}+\frac{1}{2m}\partial _{C}\left(m\omega ^{2}\right)Q^{2}
	+\frac{m'^{2}}{2m^{2}}\dot{C}Q\dot{Q}-\frac{m'^{3}}{8m^{3}}\dot{C}^{2}Q^{2}=0
	\end{align}
	which exactly coincides with the back-reaction equation for $q$.

	Finally the momentum associated with the original Lagrangian takes the following forms,
	\begin{align}
	p_{C}=\frac{\partial L_{1}}{\partial \dot{C}}=\mathcal{M}\dot{C};\qquad p_{q}=\frac{\partial L_{1}}{\partial \dot{q}}=m\dot{q}
	\end{align}
	while the momentum associated with the transformed system becomes, 
	\begin{align}
	P_{C}=\frac{\partial L_{2}}{\partial \dot{C}}=\left(\mathcal{M}-\frac{m'^{2}}{4m^{2}}Q^{2}+\frac{m''}{2m}Q^{2}\right)\dot{C};\qquad 
	P_{Q}=\frac{\partial L_{2}}{\partial \dot{Q}}=\dot{Q}
	\end{align}
	Thus the Hamiltonian in the first case reads,
	\begin{align}
	H_{1}=\frac{p_{C}^{2}}{2\mathcal{M}}+\mathcal{M}V(C)+\frac{p_{q}^{2}}{2m}+\frac{1}{2}m\omega ^{2}q^{2}
	\end{align}
	while that in the transformed system becomes,
	\begin{align}
	H_{2}&=\frac{P_{C}^{2}}{2\left(\mathcal{M}-\frac{m'^{2}}{4m^{2}}Q^{2}+\frac{m''}{2m}Q^{2}\right)^{2}}
	\left(\mathcal{M}+\frac{m'^{2}}{4m^{2}}Q^{2}\right)+\mathcal{M}V(C)
	\nonumber
	\\
	&+\frac{1}{2}P_{Q}^{2}+\frac{1}{2}\omega ^{2}Q^{2}-\frac{m'}{2m}\frac{QP_{C}P_{Q}}{\left(\mathcal{M}-\frac{m'^{2}}{4m^{2}}Q^{2}+\frac{m''}{2m}Q^{2}\right)}
	\end{align}
	Using the transformation properties:
	\begin{align}
	\frac{p_{C}}{\mathcal{M}}=\frac{P_{C}}{\left(\mathcal{M}-\frac{m'^{2}}{4m^{2}}Q^{2}+\frac{m''}{2m}Q^{2}\right)};\qquad 
	P_{Q}=\frac{p_{q}}{\sqrt{m}}+\frac{m'}{2\sqrt{m}}\frac{qp_{C}}{\mathcal{M}}
	\end{align}
	one can show that $H_{1}=H_{2}$.
	
	For the study the $Q-C$ system, we can use yet another Lagrangian $L_3$ which is numerically same as $L_1$ but functionally different. This is obtained by transforming the dynamical but without dropping the total time derivative:
	\begin{align}
	L_3(C,\dot{C},Q,\dot{Q})=\frac{1}{2}\mathcal{M}\dot{C}^2-\mathcal{M}V'(C)+\frac{1}{2}\dot{Q}^2-\frac{1}{2}\omega^2(C)+\frac{\dot{C}^2Q^2m'^2}{8m^2}-\frac{\dot{C}Q\dot{Q}m'}{2m}
	\end{align}
	The canonical momenta are given by
	\begin{align}
	\bar{P}_{C}=\partial_{C}L_3=p_{C}-\frac{m'}{2m}qp_{q}\\
	\bar{P}_{Q}=\frac{p_q}{\sqrt{m}}
	\end{align}
	These set of equations, along with the transformation equation $Q=\sqrt{m}q$ constitute a canonical transformation, which can be generated by a function $G(\bar{P}_C,\bar{P}_Q,q,C)$ given by
	\begin{align}
	G(\bar{P}_C,\bar{P}_Q,q,C)=C\bar{P}_C+q\bar{P}_Q\sqrt{m}
	\end{align}
	It can be easily verified that, the following equations are equivalent to the canonical transformation connecting $C-Q$ and $C-q$ systems.
	\begin{align}
	\partial_{q}G=p_q&&\partial_{C}G=C\\
	\partial_{\bar{P}_Q}G=Q&&\partial_{\bar{P}_C}G=C
	\end{align} 
	The Lagrangians $L_3$ and $L_1$ should be related, in general, by $L_1=L_3+dF/dt$ where, $F=-Q\bar{P}_Q-C\bar{P}_{C}C+G$. Using explicit expression for $G$ we can see that $F=0$, ensuring that $L_1=L_3$ as expected.
	The new Hamiltonian constructed out of $L_3$ is given by
	\begin{align}
	H_3=\frac{\bar{P}_C^2}{2\mathcal{M}}+\mathcal{M}V(C)+\frac{\bar{P}_Q^2}{2}+\frac{1}{2}\omega^2(C)Q^2+\frac{\bar{P}_C\bar{P}_QQm'}{2\mathcal{M}m}+\frac{\bar{P}_Q^2Q^2m'^2}{8m\mathcal{M}}
	\end{align}
	Since the generating function $G$ does not have explicit time dependence we find that $H_3$ and $H_1$ are numerically equal. Therefore, we get $H_3=H_2=H_1$(numerically).
	
	\bibliography{OscillatorRepresentations}
	
	\bibliographystyle{utphys1}
	
\end{document}